

\def\nofirstpagenoten{\nopagenumbers\footline={\ifnum\pageno>1\tenrm
\hss\folio\hss\fi}}
\def\nofirstpagenotwelve{\nopagenumbers\footline={\ifnum\pageno>1\twelverm
\hss\folio\hss\fi}}
\def\nofirstpagenotwelver{\nopagenumbers\footline={\ifnum\pageno>0\twelverm
\hss\folio\hss\fi}}

\def\leaderfill{\leaders\hbox to 1em{\hss.\hss}\hfill}

\font\rml=cmr12 scaled \magstep1  
\font\bfl=cmbx12 scaled \magstep1 
\font\itl=cmti12 scaled \magstep1 

\parindent=20pt
\def\narrow{\advance\leftskip by 40pt \advance\rightskip by 40pt}

\def\AB{\bigskip
        \centerline{\bf ABSTRACT}\medskip\narrow}
\def\nonarrower{\advance\leftskip by -40pt\advance\rightskip by -40pt}
\def\AE{\bigskip\nonarrower}

\def\boxit#1{\vbox{\hrule\hbox{\vrule\kern3pt
        \vbox{\kern3pt#1\kern3pt}\kern3pt\vrule}\hrule}}

\def\gtorder{\mathrel{\raise.3ex\hbox{$>$}\mkern-14mu
             \lower0.6ex\hbox{$\sim$}}}
\def\ltorder{\mathrel{\raise.3ex\hbox{$<$}|mkern-14mu
             \lower0.6ex\hbox{\sim$}}}
\def\dalemb#1#2{{\vbox{\hrule height .#2pt
        \hbox{\vrule width.#2pt height#1pt \kern#1pt
                \vrule width.#2pt}
        \hrule height.#2pt}}}

\font\twelvett=cmtt12 \font\twelvebf=cmbx12
\font\twelverm=cmr12 \font\twelvei=cmmi12 \font\twelvess=cmss12
\font\twelvesy=cmsy10 scaled \magstep1 \font\twelvesl=cmsl12
\font\twelveex=cmex10 scaled \magstep1 \font\twelveit=cmti12
\font\tenss=cmss10
 
 \font\ninebf=cmbx9
\font\ninerm=cmr9 \font\ninei=cmmi9
\font\ninesy=cmsy9 
\font\eightrm=cmr8
\catcode`@=11
\newskip\ttglue
\newfam\ssfam

\def\twelvepoint{\def\rm{\fam0\twelverm}
\textfont0=\twelverm \scriptfont0=\ninerm \scriptscriptfont0=\sevenrm
\textfont1=\twelvei \scriptfont1=\ninei \scriptscriptfont1=\seveni
\textfont2=\twelvesy \scriptfont2=\ninesy \scriptscriptfont2=\sevensy
\textfont3=\twelveex \scriptfont3=\twelveex \scriptscriptfont3=\twelveex
\def\it{\fam\itfam\twelveit} \textfont\itfam=\twelveit
\def\sl{\fam\slfam\twelvesl} \textfont\slfam=\twelvesl
\def\bf{\fam\bffam\twelvebf} \textfont\bffam=\twelvebf
\scriptfont\bffam=\ninebf \scriptscriptfont\bffam=\sevenbf
\def\tt{\fam\ttfam\twelvett} \textfont\ttfam=\twelvett
\def\ss{\fam\ssfam\twelvess} \textfont\ssfam=\twelvess
\tt \ttglue=.5em plus .25em minus .15em
\normalbaselineskip=14pt
\abovedisplayskip=14pt plus 3pt minus 10pt
\belowdisplayskip=14pt plus 3pt minus 10pt
\abovedisplayshortskip=0pt plus 3pt
\belowdisplayshortskip=8pt plus 3pt minus 5pt
\parskip=3pt plus 1.5pt
\setbox\strutbox=\hbox{\vrule height10pt depth4pt width0pt}
\let\sc=\ninerm
\let\big=\twelvebig \normalbaselines\rm}
\def\twelvebig#1{{\hbox{$\left#1\vbox to10pt{}\right.\n@space$}}}

\def\tenpoint{\def\rm{\fam0\tenrm}
\textfont0=\tenrm \scriptfont0=\sevenrm \scriptscriptfont0=\fiverm
\textfont1=\teni \scriptfont1=\seveni \scriptscriptfont1=\fivei
\textfont2=\tensy \scriptfont2=\sevensy \scriptscriptfont2=\fivesy
\textfont3=\tenex \scriptfont3=\tenex \scriptscriptfont3=\tenex
\def\it{\fam\itfam\tenit} \textfont\itfam=\tenit
\def\sl{\fam\slfam\tensl} \textfont\slfam=\tensl
\def\bf{\fam\bffam\tenbf} \textfont\bffam=\tenbf
\scriptfont\bffam=\sevenbf \scriptscriptfont\bffam=\fivebf
\def\tt{\fam\ttfam\tentt} \textfont\ttfam=\tentt
\def\ss{\fam\ssfam\tenss} \textfont\ssfam=\tenss
\tt \ttglue=.5em plus .25em minus .15em
\normalbaselineskip=12pt
\abovedisplayskip=12pt plus 3pt minus 9pt
\belowdisplayskip=12pt plus 3pt minus 9pt
\abovedisplayshortskip=0pt plus 3pt
\belowdisplayshortskip=7pt plus 3pt minus 4pt
\parskip=0.0pt plus 1.0pt
\setbox\strutbox=\hbox{\vrule height8.5pt depth3.5pt width0pt}
\let\sc=\eightrm
\let\big=\tenbig \normalbaselines\rm}
\def\tenbig#1{{\hbox{$\left#1\vbox to8.5pt{}\right.\n@space$}}}
\let\rawfootnote=\footnote \def\footnote#1#2{{\rm\parskip=0pt\rawfootnote{#1}
{#2\hfill\vrule height 0pt depth 6pt width 0pt}}}

\twelvepoint \nofirstpagenotwelver
  
\def\ft#1#2{{\textstyle{{#1}\over{#2}}}}
\def\frac#1#2{{{#1}\over{#2}}}
\def\1#1{\frac1{#1}} \def\2#1{\frac2{#1}} \def\3#1{\frac3{#1}}
\def\noss{\noalign{\smallskip}}
\def\sb#1{\lower.4ex\hbox{${}_{#1}$}}

\def\pa{\partial}

\def\.{\,\,,\,\,}
\def\'{\mkern 1mu}
\def\FF#1#2#3#4#5{\,\sb{#1}F\sb{\!#2}\!\left[\,{{#3}\atop{#4}}\,;{#5}\,\right]}

\def\cramp{\medmuskip = 2mu plus 1mu minus 2mu}

\def\crampest{\medmuskip = 1mu plus 1mu minus 1mu}

\cramp
\def\ll{\relax{\langle\kern-.20em \langle}}
\def\rr{\relax{\rangle\kern-.20em \rangle}}
\def\ax#1{\nwarrow\kern-1.15em #1}

\def\two{I\kern-.18em I}
\def\three{I\kern-.20em I\kern-.20em I}
\def\four{I\kern-.18em V}

\font\big=cmbx12 scaled\magstep2
\hsize=16.5 truecm
\vsize=23.0 truecm
\baselineskip 20pt
\parskip 0pt

\def\lagr{{\cal L}}
\def\noss{\noalign{\smallskip}}
\def\noms{\noalign{\medskip}}


\pageno=0
\rightline{CERN-TH.6404/92}
\rightline{February 1992}
\vskip 2truecm

\bigskip
\bigskip
\centerline{\bfl W-INFINITY AND STRING THEORY\big}

\vskip 2truecm
\centerline{\rml X. SHEN}
\bigskip
\centerline{\itl Theory Division, CERN}
\medskip
\centerline{\itl CH-1211 Geneva 23, Switzerland\big}

\vskip 2truecm

\AB
We review some recent developments in the theory of $W_\infty$. We
comment on its relevance to lower-dimensional string theory.

\AE
 \vskip 3truecm
 \hfill\break
\vskip 1truecm
\noindent CERN-TH.6404/92  \hfill\break
\noindent February 1992
\vfill
\eject

\def\slinfty{$SL(\infty,R)$\ }

\noindent
{\twelverm 1. INTRODUCTION}
\medskip

The advent of modern string theories in recent years has brought forth
a tremendous amount of investigations in two-dimensional physics.
In particular a great deal of progress has been made in two-dimensional
conformal field theories (CFTs) [1,2,3]. These theories possess the
celebrated Virasoro algebra as their underlying symmetry.

Virasoro symmetry has been well studied in both physics and mathematics
literature. When the central charge $c$ vanishes, it is essentially
the diffeomorphism group on a circle. The central extension furnishes
the algebra with many non-trivial features such as its complicated space
of representations. It is well known that for the case when $0<c<1$, a full
classification of the unitary representations of Virasoro has been given [4].
In general for the case when $c>1$, there exists no complete classification
for the unitary representations.

The reason why there exists such a classification of the unitary
representations of the Virasoro algebra with $0<c<1$ is due, to a large
extent, to the tight structure imposed by the symmetry itself. In
particular, all unitary theories with $0<c<1$ fall into the so-called
minimal models of the Virasoro algebra, whose simplicity is a non-trivial
feature of Virasoro symmetry with its central charge in such a range.
When $c\ge1$, such a concept of minimality is lost in general, and the
space of representations of Virasoro is much more complex. One way to
attack this problem is to introduce a larger symmetry, which contains
Virasoro as its subalgebra, so that a new concept of minimality with
respect to the larger symmetry emerges. These symmetries are generically
referred to as extended conformal symmetries. Theories that possess
such symmetries are referred to as the Rational Conformal Field Theories
(RCFTs) [3].

Virasoro symmetry is generated by the spin-2 stress tensor field $T(z)$.
Extended conformal symmetries include additional generators. For
example, in the case of the superconformal algebras, there exist
additional fermionic fields with appropriate conformal
spins, plus additional
bosonic fields necessary to form supermultiplets.
Another class of extended conformal algebras is the famous $W$-algebras [5,6],
the first of which, the $W_3$ algebra, was discovered by Zamolodchikov
[5]. A characteristic of this class of algebras is that they contain
fields with integral higher-spin. In this terminology, Virasoro
may be referred to as the $W_2$ algebra. In general, the existence of
additional generators indeed invokes more refined notions of minimality
so as to render the space of representations more manageable.

The motivation discussed above to search and study extended conformal
symmetries evolves around the quest to understand and classify the
two-dimensional CFTs, whose applications range from two-dimensional
critical phenomena to string theory as a candidate for the unifying theory.
{}From the
viewpoint of string theory, there is also a need to understand
symmetries other than the Virasoro algebra. In general, the underlying
symmetry structure of string theory is not known, which makes a more
coherent formulation of the theory still inaccessible. Since a symmetry on
the world-sheet of the string is often reflected in the target space-time,
investigations into algebraic structures on the two-dimensional world-sheet
may shed light on the symmetry structure of the target space-time of the
string. This strategy in
understanding string theory differs from that of string field theory.
Nonetheless it may
be a useful way to probe string theory and lead to
a successful formulation of string field theory.

Recently in the study of lower-dimensional string theories, and in
the $c=1$ (bosonic) model in particular, there emerges evidence
for the existence of elegant symmetry structures [7,8,9,10]. These
developements
have made it more imperative to better understand
symmetries larger than the Virasoro algebra.

One of the symmetries that has emerged in these investigations into
lower-dimensional string theories is some $W_\infty$ symmetry.
{}From the viewpoint of the world-sheet, on which conformal symmetry
(Virasoro) plays an important role, $W_\infty$ symmetry can be viewed
as the $N\to\infty$ limit of the extended conformal algebra $W_N$.
It is important to emphasise that such a viewpoint does not necessarily
imply that the emerging $W_\infty$-like symmetry exists on the world-sheet.
Recent analysis seems to suggest that it should more likely be a symmetry
in the configuration space of the theory.

In view of the above motivations, we shall review some recent developments
in the theory of $W_\infty$ with an eye towards its application in
studying string theories. Some of the topics of $W_\infty$ theory are
interesting in their own right, and potentially
may become relevant to string theory. They include topics such as
$W_\infty$ gravity as a higher-spin extension of ordinary
two-dimensional gravity, the $W_\infty$ string as an extension of ordinary
string theory, and the concept of universal $W$-algebra that encompasses
all finite-$N$ $W_N$ algebras. Hopefully such an effort may be of use
in making some technology developed more available.

The paper is organized as follows. In Section 2 we shall first give
the algebraic structures of various $W_\infty$ algebras, and make comments on
issues pertaining to these algebras, such as their relationships to
the area-preserving diffeomorphism of a two-surface, their subalgebras,
their relationship to other algebras {\it e.g.} the algebra of
differential operators of arbitrary degree on a circle.

Sections 3,4,5,6 and 7 are about the field theory of $W_\infty$. We shall
start in
Section 3 with some known global realizations of $W_\infty$. Section 4
covers the
classical formulation of $W_\infty$ gravity and the $W_\infty$ string,
where the
introduction of $W_\infty$ gauge fields make the symmetry locally realized.
Sections
5,6 and 7 cover three different topics in the quantization of $W_\infty$
gravity and
the $W_\infty$ string. In the first topic, we demonstrate that a classical
$w_\infty$
gravity model is quantum mechanically inconsistent and deforms into a
quantum $W_\infty$
gravity upon quantization. The second topic concerns the BRST analysis of
$W_\infty$. The third contains the work that demonstrates the existence
of an
$SL(\infty,R)$ Kac-Moody symmetry in $W_\infty$ gauge theories.

In Section 8, we shall look into the recent investigation into the
lower-dimensional
string models, and draw some parallels between the symmetries in these
models and the
$W_\infty$ symmetry discussed in the previous sections. We summarize
the paper
in Section 9, and make brief reference to other topics left out of this
review in the
field of $W$ gravity and the $W$ string.

\bigskip
\bigskip
\noindent
{\twelverm 2. THE ALGEBRAIC STRUCTURE OF THE $W_\infty$ ALGEBRAS}
\medskip

The $W_\infty$ algebras are bosonic extensions of the Virasoro
algebra. They can be viewed as the $N\to\infty$ limits of finite-$N$
$W_N$ that contains generators of conformal spin 2,3,$\ldots$,N.
Thus a generic $W_\infty$ algebra contains an infinite number of
generating currents of conformal-spin 3,4,$\ldots$,$\infty$, in
addition to the spin-2 stress tensor of Virasoro.

Since the procedure of taking $N\to\infty$ limit is rather subtle,
it is believed that there exist more than one $W_\infty$ algebra
that correspond to the same $W_N$ algebra.
In particular there is a non-linear $W_\infty$ algebra [11] as a
limit as well as the linear $W_\infty$ algebras that have been discovered
[12,13,14].
In this paper we shall only be concerned with the linear versions
of $W_\infty$.

A striking feature of the linear $W_\infty$ algebra is its resemblance
to Virasoro, which will become more and more evident as we move on.
Essentially what can be achieved in the case of Virasoro can be applied
and generalized rather straightforwardly to the case of $W_\infty$.
Since it is much less demanding technically to work with Virasoro
than $W_\infty$, we shall be content as often as possible to illustrate
basic ideas in the case of Virasoro, and then give the answers
for the case of $W_\infty$.

Let us first start with the Virasoro algebra.
In its Fourier modes $L_n$ of the spin-2 stress tensor field $T(z)$
given by
$$
T(z)\equiv \sum{L_n\over z^{n+2}}, \eqno(1)
$$
the algebra reads
$$
[L_m,L_n] = (m-n) L_{m+n} + {c\over 12} (m-1)m(m+1) \delta_{m+n,0}.
\eqno(2)
$$
In terms of the Operator Product Expansion (OPE) of the stress tensor
$T(z)$, the algebra is given by
$$
T(z) T(w)\sim {\pa T(w) \over z-w} + { 2 T(w)\over (z-w)^2} +
{c/2\over (z-w)^4}.\eqno(3)
$$

Note that there exists an $SL(2, R)$ subalgebra formed by $\{L_{-1},
L_0, L_{-1}\}$. As will become clear presently when the $W_\infty$
algebras are given, the covariance property under this subalgebra
of Virasoro dictates their algebraic structures to a large extent.

The centerless Virasoro algebra can be viewed as the algebra of a vector
differential on a circle parametrized by $\theta$ in the following
way:
$$
L_m = e^{im\theta}{d\over d\theta},\eqno(4)
$$
which generates the diffeomorphism group $Diff S^1$. From this
viewpoint, Virasoro is the centrally extended algebra of $Diff S^1$.
This observation finds its natural generalization to the case of
$W_\infty$.

There are many ways to enlarge the Virasoro algebra to various extended
conformal algebras. Among them is the famous $W_3$ algebra of
Zamolodchikov [5], in which there exists a spin-3 generator $W(z)$
in addition to the spin-2 $T(z)$. This can be generalized to the
finite-$N$ $W_N$ algebra containing fields with spin $2,3,\cdots,N$ [6],
and their supersymmetric generalizations [7]. The unique
characteristic of these algebras, which makes them very interesting
but technically very difficult to work with, is the non-linearity
inherent in the structure, due to the introduction of higher-spin
generators. For example, in the case of $W_3$, apart from the OPE
between $T$ and $W$ given as
$$
T(z) W(w)\sim {\pa W(w) \over z-w} + {3 W(w)\over (z-w)^2},\eqno(5)
$$
the OPE between $W$ reads
$$
\eqalign{
W(z) W(w)&\sim {c/3\over(z-w)^6} + {16\over 22+5c} \Big({2\Lambda\over
(z-w)^2} + {\pa\Lambda\over z-w}\Big)\cr
&+\ft1{15}{\pa^3 T\over z-w}+\ft3{10}{\pa^2 T\over (z-w)^2}
+{\pa T\over (z-w)^3} +{2T\over (z-w)^4}\cr}
\eqno(6)
$$
where the $\Lambda$ field is defined by
$$
\eqalign{
\Lambda(z)&\equiv \big(TT\big)-\ft3{10}{\pa}^2 T \cr
\noms
\big(TT\big)(z) &\equiv \oint {dw \over 2\pi i}{T(w)T(z)\over w-z}\cr}.
\eqno(7)
$$
The appearance of this composite spin-4 field $\Lambda$ is the source of
non-linearity. However this new feature also
drastically increases the magnitude of difficulty in {\it e.g.}
formulating field theory exhibiting such a symmetry at both
classical and quantum levels. In fact, for arbitrary $N$, due to
the complexity arising from the non-linearity, the structure
constants of $W_N$ are in general not known explicitly.

In order to circumvent the situation of having to deal with
non-linearity, it is conceivable that if the composite higher-spin
terms such as the $\Lambda(z)$ in $W_3$, which is necessary for the
closure, are replaced by new fundamental fields,
non-linear terms may disappear. The consequence is that one may need
to introduce more and more fundamental higher-spin fields.
Thus it may only be possible once one has introduced sufficiently
many independent fields. A reasonable set of fields, for example,
includes one field for each spin $s\ge 2$. Indeed, as has been shown
in [12,13], such a choice of field content does yield a consistent
algebra, which preserves most of the important features of Virasoro and
$W_N$, such as non-trivial central extensions for higher-spin
generators. This algebra is thus naturally called the $W_\infty$
algebra.

Historically the original discovery of the $W_\infty$ algebras
employed a method in which a particularly natural form for the
sought-after algebra was assumed and then the structure constants
were calculated by imposing Jacobi identities. Apart from the
naturalness of the assumptions that went into the construction, it
was quite a miracle that solutions were found at all.
It has only become clear later by considerations from other angles
that there must exist algebras as such. We shall make remarks to
make this point more explicit in the relevant context.

The {\it ansatz} in [12,13] is of the following form:
$$
[V^i_m,V^j_n]=\sum_{\ell\ge0} g^{ij}_{2\ell}(m,n)V^{i+j-2\ell}_{m+n}
+c_i(m)\delta^{ij}\delta_{m+n,0},\eqno(8)
$$
where $V^i_m$ denotes the $m$'th Fourier mode of a spin-$(i+2)$ field,
and the central terms are assumed to take the form of that of the $W_N$
algebras given by
$$
c_i(m)=(m^2-1)(m^2-4)\cdots(m^2-(i+1)^2)c_i,\eqno(9)
$$
The solutions to the Jacobi identities are thus given by
$$
g_{\ell}^{ij}(m,n)=\frac{\phi^{ij}_{\ell}}{2(\ell+1)!}
\,N^{ij}_{\ell}(m,n),\eqno(10)
$$
where the $N^{ij}_{\ell}(m,n)$ are given by
\crampest
$$
N_{\ell}^{ij}(m,n)
=\sum_{k=0}^{\ell+1}(-1)^k{\ell+1\choose k}[i+1+m]_{\ell+1-k}
[i+1-m]_k[j+1+n]_k[j+1-n]_{\ell+1-k}.
\eqno(11)
$$
In (11), $[a]_n$ denotes the descending Pochhammer symbol given by
$$
[a]_n\equiv
a(a-1)\cdots (a-n+1)=a!/(a-n)!\eqno(12)
$$
The functions $\phi^{ij}_{\ell}$ can be expressed as
$$
\phi^{ij}_{\ell} =
\FF43{-\ft12\.\ft32\.-\ell-\ft12\.-\ell}{-i-\ft12\.-j-\ft12\.i+j-2\ell+\ft52}1
,\eqno(13)
$$
where the right-hand side is a
Saalsch\"utzian $_4F_3(1)$ generalized hypergeometric function [16].
The central charges $c_i$ are given by
$$
c_i={2^{2i-3}i!(i+2)!\over(2i+1)!!(2i+3)!!}c.\eqno(14)
$$

\bigskip\bigskip
\noindent
{\twelverm REMARKS}:
\medskip
\noindent
* \ {\twelvesl $SL(2,R)$ Covariance and ``Wedge'' Algebra\big}
\medskip

The functions $N^{ij}_{2\ell}(m,n)$  are related to the Clebsch-Gordan
coefficients of $SU(2)$ or $SL(2,R)$, while the functions
$\phi^{ij}_{2\ell}$ are related (in a formal sense, at least) to Wigner
6-$j$ symbols.  These facts are indicative of an underlying $SL(2,R)$
structure of the $W_\infty$ algebra, and indeed this is the case.
As indicated previously for Virasoro, the generators $L_{-1}$,
$L_0$ and $L_1$ give an anomaly-free $SL(2,R)$ subalgebra. Since the
Virasoro algebra is itself a subalgebra of $W_\infty$, this implies that
we also have an $SL(2,R)$ subalgebra in $W_\infty$, generated by
$V_{-1}$, $V^0_0$ and $V^0_1$.  This $SL(2,R)$ in fact forms the bottom
``rung'' of a wedge of generators, comprising the $V^i_m$ with $-i-1\le
m\le i+1$ for a given value of $i$; the generators on that rung transform
as the $(2i+3)$-dimensional representation of $SL(2,R)$.  The set of all
wedge generators, which we shall call the ``wedge'' algebra,
give rise to an $SL(\infty,R)$ algebra [13].

\medskip
\noindent
* \ {\twelvesl The Wedge Algebra as Tensor algebra of $SL(2,R)$\big}
\medskip

The $SL(\infty,R)$ wedge subalgebra of $W_\infty$ can be constructed
as a tensor algebra of $SL(2,R)$, modded out by the ideal generated by
$C_2-s(s+1)$, where $C_2$ is the Casimir operator of $SL(2,R)$, and $s$ is
a constant that must be chosen to be zero in this case [17].
One obtains inequivalent \slinfty algebras by taking
values other than $s(s+1)=0$ for the quadratic Casimir. Specifically
the generators $V^i_m$ with $-1-i\le m \le i+1$, transforming
as the $(2i+3)$-dimensional representation of $SL(2,R)$, are constructed
from appropriate polynomials of degree $i+1$ in the generators of
$SL(2,R)$. If one starts with $V^i_{i+1}\equiv (V^0_1)^{i+1}$, then
constructs the generators in the same representation of $SL(2,R)$ by
acting on $V^i_{i+1}$ with the lowering operator $V^0_{-1}$,
modulo the identification imposed by the ideal given above, one
obtains the entire \slinfty algebra.
This family of \slinfty algebras parametrized by
$s$ will be explicitly given in Eqs.(118-120), when they are an important
part of the
discussion on the existence of \slinfty Kac-Moody symmetry in quantum
$W_\infty$ gravity.

\medskip
\noindent
* \ {\twelvesl The Tensor Algebras and Area-preserving Diffeomorphism\big}
\medskip

There exists an intriguing relationship between this type of tensor
algebras and the algebra of an area-preserving diffeomorphism on a
two-dimensional surface [18,19], first discovered in [18]. Here we
only sketch the basic idea of such an identification.

Consider the
area-preserving algebra for the two-dimensional sphere, which can
be viewed as being embedded in a three-dimensional Euclidean
space with coordinates $x^i, i=1,2,3,$ defined by the following
constraint
$$
x^i x^i = r^2, \eqno(15)
$$
where $r$ is a constant.
Let us introduce a Lie bracket given by
$$
\{x^i, x^j\} = \epsilon_{ijk} x^k,\eqno(16)
$$
so that for functions $A(x)$ and $B(x)$ on the sphere the Lie bracket
takes the form
$$
\{A(x), B(x)\} = \epsilon_{ijk} x^i\pa_j A(x)\pa_k B(x). \eqno(17)
$$
The transformation law on $x^i$ generated by the function $f$ is given by
$$
\delta_f x^i =\epsilon_{ijk} x_j {\pa f\over\pa x_k}.\eqno(18)
$$
It is easy to check that the set of all functions generates the
algebra of area-preserving diffeomorphism of the sphere.

In order to identify this algebra with a specific tensor algebra
discussed above, one chooses the spherical harmonics given by
$$
Y_{1,1}\sim x_1+ix_2,\quad Y_{1,0}\sim x_3,\quad Y_{1,-1}\sim x_1-ix_2,
\eqno(19)
$$
from which one can construct the basis for polynomials of higher degree
by taking $Y_{\ell,\ell}$ to be $(Y_{1,1})^\ell$ and acting on
$Y_{\ell,\ell}$ with $Y_{1,-1}$ in the sense of the above Lie bracket.
Numerologically one sees that if the following identification
$$
V^0_1\to Y_{1,1},\quad V^0_0\to Y_{1,0},\quad V^0_{-1}\to Y_{1,-1}
\eqno(20)
$$
is made, then there exists a one-to-one correspondence between the
generators of the tensor algebra and polynomials in the basis
for generating the area-preserving diffeomorphism on the sphere.
Indeed, it can be made rigorous that the area-preserving
diffeomorphism on the two-dimensional sphere is isomorphic to
the $SL(\infty,R)$ with the parameter $s=\infty$ [19].

Such an identification lends a geometrical flavor to the wedge
algebra of $W_\infty$ ($s=0$), or more precisely
to the $SL(\infty,R)$ algebra with $s=\infty$.
However, the full $W_\infty$ algebra with its Fourier modes ``extending
beyond the wedge'' does not share this link to geometry. Although there
are interesting and tantalizing ideas on how to establish such a link,
none of them seems to be solidly demonstrated at this point. Thus
$W_\infty$ is different from the algebra of area-preserving
diffeomorphism on a two-dimensional surface, and is certainly by no means
included in the class of area-preserving diffeomorphism algebras.

\medskip
\noindent
* \ {\twelvesl The $w_\infty$ Algebra and Area-preserving Diffeomorphism
\big}
\medskip

An alternative link to geometry that has been demonstrated is the
following. If we perform the rescaling
$V^i_m\to q^{-i}V^i_m$ for $W_\infty$, then one can easily
see that in the limit $q\to 0$, all the lower-order terms on the
right-hand side of (3) disappear, as do all the central terms apart from
the one in the Virasoro subsector. So we
see that in this limit the $W_\infty$ algebra contracts down to
an algebra with the same content of generators, which is commonly
referred to as the $w_\infty$ algebra, first discovered in [20].
It can be written in the following simple form
$$
[v^i_m,v^j_n]=\Big((j+1)m-(i+1)n\Big)v^{i+j}_{m+n},\eqno(21)
$$
where we use $v^i_m$ to denote the generators so as to distinguish them
from the $V^i_m$ of $W_\infty$.

Now the $w_\infty$ algebra (21) can be enlarged to $w_{1+\infty}$,
with conformal spins $s=i+2\ge1$ simply by allowing the indices $i$ and
$j$ to take the value $-1$ as well as the non-negative integers. The
resulting algebra admits a geometrical interpretation [20,21,22], as the
algebra of smooth symplectic ({\it i.e.} area-preserving)
diffeomorphisms of a cylinder [22]. To see this,
consider the  set of functions $u^\ell_m=-i e^{mx}y^{\ell+1}$ on
a cylinder $S^1\times R$, with coordinates $0\le x\le 2\pi$, $-\infty \le
y \le \infty$.  These functions form a complete set if $-\infty \le m\le
\infty$ and $\ell\ge-1$.  The symplectic transformations of the cylinder,
which preserve the area element $dx\wedge dy$, are
generated by $\delta x^\mu=\{\Lambda,x^\mu\}$, where $\Lambda$ is an
arbitrary function and $\{f,g\}$ is the Poisson bracket
$$
\{f,g\}={\partial f\over\partial x}{\partial g\over\partial y}-
{\partial f\over\partial y}{\partial g\over\partial x}.\eqno(22)
$$
One can easily see, by expanding $\Lambda$ in terms of the
the basis $u^\ell_m$, that the
commutator of symplectic transformations satisfies precisely the
algebra (1), with $v^i_m\to u^i_m$ and $[f,g]\to \{f,g\}$.

So far we have illustrated two different ways of making contact with
the area-preserving diffeomorphism. Although both ways
employ Poisson brackets, there exists a subtle difference between them.
In the first approach, the $m$ index of $V^i_m$ is confined with the
wedge, while the $m$ index of $v^i_m$ has to cover the whole range in
order to form a complete basis of functions on a cylindrical surface.
This difference becomes important in the context of the $c=1$ string
coupled
to two-dimensional gravity, where there have been discussions on
algebras similar to the area-preserving diffeomorphism on two-dimensional
surface.

\medskip\noindent
* \ {\twelvesl The $W_{1+\infty}$ Algebra\big}
\medskip

Having extended $w_\infty$ to include a spin-1 generator, we are
naturally led to consider if it is possible to implement such
an extension for the uncontracted $W_\infty$. The answer is positive [14].
Explicitly, the $W_{1+\infty}$ algebra has the same form as (8) and (9)
except that the structure constants are now
$$
\eqalign{
{\tilde g}_{\ell}^{ij}(m,n)&=\frac{{\tilde \phi}^{ij}_{\ell}}{2(\ell+1)!}
\,N^{ij}_{\ell}(m,n),\cr
{\tilde\phi}^{ij}_{\ell}& =
\FF43{\ft12\.\ft12\.-\ell-\ft12\.-\ell}{-i-\ft12\.-j-\ft12\.i+j-2\ell+\ft52}
1,\cr}\eqno(23)
$$
the $N^{ij}_\ell(m,n)$ remain the same,
and the central charges ${\tilde c}_i$ are given by
$$
{\tilde c}_i={2^{2i-2}((i+1)!)^2\over (2i+1)!!(2i+3)!!}c\eqno(24)
$$
The wedge subalgebra in this case can be enlarged to
include the single generator $V^{-1}_0$ at the apex, giving the algebra
$GL(\infty,R)$, which again can be constructed as a tensor algebra of
$SL(2,R)$ modulo the same ideal with $s=-\ft12$.
A contraction of $W_{1+\infty}$, analogous to that
described above for $W_\infty$, yields $w_{1+\infty}$ as a classical
limit.

Note that in this paper we shall
uniformly adopt the notation that structure constants, central charges,
currents, gauge fields, etc., for the case of $W_\infty$ gravity will be
denoted (as we have been doing so far) by untilded quantities, as
opposed to the tilded quantities of $W_{1+\infty}$.

Since the spin content of $W_{1+\infty}$ contains that of
$W_\infty$, with an additional spin-1 field, one might think
that  it should be possible to view $W_\infty$ as a subalgebra of
$W_{1+\infty}$. Indeed, this is the case; the details may be found in [17].
An interesting feature of this inclusion is that the central charge of
the $W_\infty$ subalgebra is $-2$ times the $W_{1+\infty}$ central
charge.

\medskip
\noindent
*\ {\twelvesl $W_\infty$ and the Algebra of All Differential Operators on
$S^1$\big}
\medskip

To close this section, we shall present another viewpoint for $W_\infty$,
which extends the well-known interpretation of the Virasoro algebra as
the centrally-extended $Diff S^1$. Intuitively it is very natural to
expect that the centerless $W_\infty$ algebra be realized in terms
of all differential operators on a circle, given the fact that the
centerless Virasoro algebra is represented by the vector differential
on a circle. This is indeed the case, as was demonstrated in [17], for the
centerless $W_\infty$ algebra. The question that remains is whether there
exists a unique central extension or cocycle structure for the
algebra of all differentials on a circle  that coincides with
the one given in the $W_\infty$ algebra. Recently an affirmative answer
has been given in [23,59].

This standpoint to view $W_\infty$ as an extension of Virasoro was
suggested some time ago [24] in the context of fermion bilinears. From
this angle, the existence of an algebra as such is very transparent. The
more difficult part is to explicitly obtain the structure constants
as well as the structure for central charges. The original brute-force
method [12] that has produced beautiful yet inexplicable algebraic
structure has been nicely complemented by the geometrical
understanding in ref.[23,59].

\bigskip
\bigskip
\noindent
{\twelverm 3. REALIZATIONS OF $W_\infty$}
\medskip

Firstly we shall summarize the known global realizations of $W_\infty$.
In comparison to the many known classes of realizations of Virasoro,
realizations of $W_\infty$ in field theory are relatively scarce,
despite the kinship that we have stressed between the two. So far, there
exist essentially only two realizations for $W_\infty$. The first
one discovered in [25] is a bosonic realization, while the second
[26,27] is a fermionic realization. A common feature shared by the two
realizations is that the generating currents of $W_\infty$ are built
out of bilinears of either a boson or a fermion. Thus the transformations
of $W_\infty$ currents on either a boson or a fermion in these two
realizations are always linear. Non-linear realizations of $W_\infty$
may appear by bosonizing either fermion [28] or boson [29].
A more detailed summary of these realizations and their interrelationships
can be
found in [29].

Consider a free, complex scalar field $\phi$, with OPE
$$
\phi^*(z)\phi(w)\sim-\log(z-w).\eqno(25)
$$
Virasoro can be realized in terms of $\phi$ simply as follows.
$$
T(z)=-\pa\phi^*\pa\phi.\eqno(26)
$$
To realize higher-spin currents, one considers bilinears of $\phi$
with higher numbers of derivatives.
One can easily establish that
at each order in the total number of derivatives distributed over the
two fields, there is exactly one independent current; any other
combination of the same number of derivatives distributed over the two
fields can always be expessed as a linear combination of the independent
current at that order together with derivatives of the lower-spin
independent currents.  Thus it is clear that the OPEs of all possible
higher-spin independent currents will form a closed algebra.
It turns out that these bilinears indeed give a realization of $W_\infty$.
{}From this viewpoint, the existence of $W_\infty$ is very transparent.
The currents $V^i(z)$ of $W_\infty$, related to the Fourier-mode
components $V^i_m$ by
$$
V^i(z)\equiv \sum_m V^i_m z^{-m-i-2},\eqno(27)
$$
are given by [25]
$$
V^i(z)=\sum_{k=0}^i \alpha^i_k:\partial^{i-k+1}\phi^*
\partial^{k+1}\phi:,\eqno(28)
$$
where the constants $\alpha^i_k$ are given by
$$
\alpha^i_k=(-)^{i+k+1}{2^{i-1}(i+2)!\over(2i+1)!!(i+1)}{i+1\choose k}
{i+1\choose k+1}.\eqno(29)
$$
This particular choice for the coefficients is the
unique possibility (up to overall $i$-dependent scalings) that ensures
that the central terms in the algebra are diagonal, {\it i.e.} that they
arise only between fields of the same spin.  Together with the specific
$i$-dependence in (29), this choice gives precisely the standard form of
the $W_\infty$ algebra.

There is a straightforward procedure for re-expressing the
Fourier-mode forms of the $W_\infty$ algebra (8) and (9)
in terms of operator-product expansions of the corresponding currents
defined by (27). To do that, we first define $f^{ij}_\ell(m,n)$
given by
$$
f_{\ell}^{ij}(m,n)=\frac{\phi^{ij}_{\ell}}{2(\ell+1)!}
\,M^{ij}_{\ell}(m,n),\eqno(30)
$$
where the $M^{ij}_{\ell}(m,n)$ are given by
\crampest
$$
M_{\ell}^{ij}(m,n)=\sum_{k=0}^{\ell+1}(-1)^k{\ell+1\choose k}(2i-\ell+2)_k
[2j+2-k]_{\ell+1-k} m^{\ell+1-k} n^k,
\eqno(31)
$$
and $\phi^{ij}_{\ell}$ is the same as given in (13). Here $[a]_n$ is the
same as
that given in (12) and $(a)_n$ is given by
$$
(a)_n=a(a+1)\cdots(a+n-1).\eqno(32)
$$
Thus in terms of the polynomials $f^{ij}_\ell(m,n)$, the $W_\infty$
algebra in
Fourier	mode form in (8) and (9) can be expressed concisely in OPE by the
following:
$$
V^i(z) V^j(w) \sim -c_i\delta^{ij}(\pa_z)^{2i+3}{1\over z-w}
-\sum_\ell f^{ij}_{2\ell}(\pa_z,\pa_w){V^{i+j-2\ell}(w)\over z-w}.\eqno(33)
$$
Now one can verify that indeed the OPEs for the
currents defined in (28) agree with those for the $W_\infty$
algebras with $c=2$.

Another characteristically similar realization of $W_\infty$ is obtained
in terms of bilinears of a free, complex fermion $\psi$, with OPE
$$
\overline{\psi(z)}\psi(w) \sim {1\over z-w}.\eqno(34)
$$
The spin-2 stress tensor $T$ is realized as
$$
T(z)=\pa\overline{\psi}\psi,\eqno(35)
$$
with $c=-2$, which is often known as a (0,1) ghost realization of
Virasoro. The higher-spin currents of $W_\infty$ are now given by
$$
V^i(z)=\sum_{k=0}^{i+1}\beta^i_k\pa^k{\overline\psi}\pa^{i+1-k}\psi,
\eqno(36)
$$
where $\beta^i_k$ is given by
$$
\beta^i_k={i+1\choose k}{(i+3-k)_k(-i)_{i+1-k}\over (i+2)_{i+1}}.\eqno(37)
$$

Since the transformations of $W_\infty$ on a generic field $\Phi$
are given by
$$
\delta_{k_i}\Phi\equiv \oint{dz\over 2\pi i} k_i(z)V^i(z)\Phi,\eqno(38)
$$
where $k_i$ are transformation parameters of the $i$-th current $V^i$,
the fields $\psi$, $\phi$ and their conjugates transform linearly,
owing to the fact that the currents of $W_\infty$ are all bilinears.
This linearity of transformation laws will be important when we quantize
field
theories with local $W_\infty$ symmetry and analyze their anomaly structures,
which will be discussed in the first topic of quantization.

A non-linear realization of $W_\infty$ is obtained by bosonizing
the fermion $\psi$, as was first done in [28]. With the identification
$\psi\to e^\varphi$ and $\overline{\psi}\to e^{-\varphi}$, and the OPE
for $\varphi$ given by
$$
\varphi(z)\varphi(w)\sim {\hbox{log}}(z-w),\eqno(39)
$$
the currents $V^i(z)$ of $W_\infty$ in terms of the scalar field $\varphi$
can be obtained by making the following replacement in (36) :
$$
 \pa^k{\overline\psi}\pa^{i+1-k}\psi\equiv\sum^{i+2}_{\ell=k+1}
{(-1)^{\ell+k+1}\over \ell}{i+1-k\choose \ell-1-k}\pa^{i+2-\ell}P^{(\ell)}(z),
\eqno(40)
$$
where $P^{(\ell)}(z)$ is given by
$$
P^{(\ell)}(z)= e^{-\varphi(z)}\pa^\ell e^{\varphi(z)}.\eqno(41)
$$

Since the currents are non-linear in the new scalar field $\varphi$,
they induce non-linear transformations on $\varphi$. For example,
the leading order in $\varphi$ of the current $V^i$ is
$\ft1{i+2}(\pa\varphi)^{i+2}$, and the corresponding transformations
on $\varphi$ resulting from such a term in the generating current are
given by
$$
\delta\varphi= k_i(\pa\varphi)^{i+1},\eqno(42)
$$
which is manifestly non-linear in $\varphi$.

One can easily check that the transformations given in (42) in fact
close to form an algebra,
$$
[\delta_{k_i},\delta_{k_j}]\varphi=\delta_{k_{i+j}}\varphi,\eqno(43)
$$
where $k_{i+j}$ is given by
$$
k_{i+j}=(j+1)k_j\pa k_i-(i+1)k_i\pa k_j.\eqno(44)
$$
One sees that this is nothing but the $w_\infty$ algebra. Algebraically
$w_\infty$ arises as a contraction of $W_\infty$, as discussed in
the previous section. In the context of field theory, the procedure of
extracting a realization of $w_\infty$ from a bilinear-fermion
realization of $W_\infty$ involves a rather subtle intermediate step
of bosonization. The leading order terms in the bosonized
field $\varphi$ generate transformations that close to form $w_\infty$.
In the language of OPEs among the currents containing only the leading
order terms in $\varphi$, the closure of $w_\infty$ is maintained
if only single contractions of the field $\varphi$ are allowed, which are
equivalent to taking the classical Poisson bracket of $\varphi$. Thus
$w_\infty$ is realized classically in the scalar field $\varphi$, with
its currents $v^i(z)$ given by
$$
v^i(z)=\ft1{i+2}(\pa\varphi)^{i+2}.\eqno(45)
$$

We shall end our discussion on realizations of $W_\infty$ with some
comments about $w_\infty$. Algebraically $w_\infty$ does not retain
one of the non-trivial properties of Virasoro, namely the central
extensions.In the last topic of  quantum $W_\infty$ gravity, we shall show
that
quantum $w_\infty$ theory seems to have little dynamics, due to the lack of
central terms in the algebra. However, owing to the simplicity in its
structure
constants, it is often much easier to find a classical realization of
$w_\infty$, such as the one given above. Having such a classical realization,
one can proceed to build a classical $w_\infty$ gauge theory that
is simple enough to illustrate the basic points, a task which we shall
undertake presently. Such a theory, in the process of quantization,
picks up additional terms in order to achieve quantum consistency in
such a way that the underlying symmetry deforms into the full
$W_\infty$. Therefore classical realizations of $w_\infty$ are
important and can serve as convenient starting points for addressing many
issues of $W_\infty$.

\bigskip
\bigskip
\noindent
{\twelverm 4. $W_\infty$ GRAVITY AND $W_\infty$ STRING}
\medskip

     Two-dimensional gravity can be thought of as a gauging of the
Virasoro algebra.  An analogous gauging of the $W$
algebras will therefore give higher-spin generalizations of
two-dimensional gravity.  Such theories are known generically as
$W$ gravity theories.  They have been discussed in the context of a
chiral gauging of $W_3$ [30]; a non-chiral gauging of $W_3$ [31]; chiral and
non-chiral $w_\infty$ [32],  $W_\infty$ and super $W_\infty$ [33].

The starting point for our $w_\infty$ gravity is a free Lagrangian for the
scalar field $\varphi$, of the form
$$
{\cal L}=\ft12\partial\varphi {\bar\partial}\varphi,\eqno(46)
$$
where $z$ and $\bar z$ are the coordinates of the two-dimensional
space-time. This action has many global symmetries.  In fact,
because of the factorization into left-moving and right-moving sectors in
two dimensions, a ``global'' symmetry typically means one that has a
parameter that depends on only $z$ or $\bar z$, but not both.

One particular global symmetry of (46) consists of transformations
given in (42), which form the $w_\infty$ algebra.
This symmetry can be made local by introducing
gauge fields $A_\ell$ for each of the spin-$(\ell+2)$ conserved currents
$(\partial\varphi)^{\ell+2}$ corresponding to the symmetries given in (42)
with parameter $k_i$ dependent on both $z$ and $\bar z$, and writing
$$
{\cal L}=\ft12(\partial\varphi {\bar\partial}\varphi) -\sum_{\ell\ge0}
{1\over \ell+2}A_\ell (\partial\varphi)^{\ell+2}.\eqno(47)
$$
One finds that this is invariant under (42) provided that the gauge fields
transform as
$$
\delta A_\ell={\bar\partial} k_\ell-\sum_{j=0}^{\ell+1}[(j+1) A_j \partial
k_{\ell-j}-(\ell-j+1)k_{\ell-j}\partial A_j].\eqno(48)
$$
In particular, if we focus attention on the spin-2 sector only, we
recover two-dimensional gravity in the chiral gauge.

     The above chiral gauging can be extended to a non-chiral one by
observing that the free action (46)  is also invariant under a second
copy of $w_\infty$, where $\partial$ in (42)  is replaced by
${\bar\partial}$, and the parameters of the transformations are taken to
depend upon $\bar z$ only.  The two copies of $w_\infty$ are made into
local symmetries by introducing two sets of gauge fields, $A_{\ell}$ and
${\bar A}_\ell$ , where the $A_{\ell}$ gauge the original copy of $w_\infty$,
and
the ${\overline{A_\ell}}$ gauge the second copy.  The action in this case
becomes more
complicated, and is most conveniently written by introducing auxiliary
fields $J$ and
$\bar J$. The required action is then given by [32]
$$
\eqalign{
{\cal L}=&-\ft12\partial\varphi{\bar\partial}\varphi-J {\bar J} +
{\bar\partial}\varphi J +\partial\varphi{\bar J} \cr
&-\sum_{\ell\ge0}\ft1{\ell+2}\Big({\bar A}_\ell {\bar J}^{\ell+2}+
A_{\ell} J^{\ell+2}\Big),\cr}\eqno(49)
$$
The equations of motion for the auxiliary fields $J$ and $\bar J$
give
$$
\eqalign{
J&=\partial\varphi-\sum_{\ell=0} {\bar A}_\ell {\bar J}^{\ell+1},\cr
{\bar J}&={\bar\partial}\varphi-\sum_{\ell=0}A_{\ell} J^{\ell+1}.\cr}
\eqno(50)
$$

It is straightforward to check that this Lagrangian is invariant under
the $k_{\ell}$ and ${\bar k}_\ell$ gauge transformations
$$
\eqalign{
\delta\varphi&=\sum_{\ell\ge-1}\Big({\bar k_\ell}{\bar J}^{\ell+1}
+k_\ell J^{\ell+1}\Big)\cr
\delta A_\ell&={\bar\partial} k_\ell -
\sum_{j=0}^{\ell +1}[(j+1)A_j\partial k_{\ell-j} -
(\ell-j+1)k_{\ell-j} \partial A_j]\cr
\delta {\bar A}_\ell&=\partial{\bar k}_\ell -
\sum_{j=0}^{\ell +1}[(j+1){\bar A}_j{\bar\partial} {\bar k}_{\ell-j} -
(\ell-j+1){\bar k}_{\ell-j} {\bar\partial} {\bar A}_j]\cr
\delta J&=\sum_{\ell\ge-1}\partial(k_\ell J^{\ell+1})\cr
\delta{\bar J}&=\sum_{\ell\ge-1}{\bar\partial}({\bar k}_\ell
{\bar J}^{\ell+1}).
\cr}
\eqno(51)
 $$
Note that $J$ is inert under the ${\bar k}$
transformations while $\bar J$ is inert under $k$ transformations.

For $W_\infty$, one proceeds in a similar manner to the one described
above, starting, for example, from a free Lagrangian of a complex
fermion $\psi$ given by
$$
{\cal L}=\overline{\psi}{\bar\pa}\psi. \eqno(52)
$$
One sees that there exist globally conserved currents of $W_\infty$ given
in (36), which induce the following global transformations on $\psi$
and $\overline{\psi}$ according to (38):
$$
\eqalign{
\delta_{k_i}\psi&=\sum_{k=0}^{i+1}(-1)^{k+1}\beta^i_k\pa^k(k_i\pa^{i+1-k}\psi)
\cr
\delta_{k_i}\overline{\psi}&=\sum_{k=0}^{i+1}(-1)^{i+1-k}\beta^i_k
\pa^{i+1-k}(k_i\pa^k{\overline\psi}) \cr}\eqno(53)
$$

To gauge the chiral $W_\infty$ symmetry , we now allow the parameters $k_i$
to depend on $\bar z$ as well as $z$. Now the free action ${\cal L}$ is not
invariant, and the remaining term in the variation of ${\cal L}$ arising from
the local parameters $k_i(z,{\bar z})$ reads
$$
\delta_{k_i}{\cal L}=-{\bar\pa}k_i\sum_{k=0}^{i+1}\beta^i_k
\pa^k{\overline\psi}\pa^{i+1-k}\psi=-{\bar\pa}k_iV^i(z),\eqno(54)
$$
so we must also introduce gauge fields
$A_i$ and add the (gauge field)-(current) coupling terms to the Lagrangian:
$$
 \lagr= \overline{\psi} \pa_- \psi + A_i V^i. \eqno(55)
$$
The Noether procedure now goes as follows.
Now the variation in (54) is cancelled by the leading-order transformation
of the gauge field
which is of the form $\delta A_i = {\bar\partial} k_i +\cdots$. We next vary
the
currents by using the formula (38) with $\Phi$ taken to be $V^j$.
The result can conveniently be expressed in the form
$$
\delta_{k_j}V^i(z)=\sum_\ell \int dw k_j(w)f^{ij}_{2\ell}(\partial
_z, \partial_w)\Big(\delta(z-w) V^{i+j-2\ell}\Big),\eqno(56)
$$
where $f^{ij}_{2\ell}(m,n)$ is given in (30) .We see that this variation
is cancelled by adding terms to the transformation rule of the gauge field
$A_i$ so that its total variation $\delta A_i={\bar\partial} k_i +\hat\delta
A_i$ is given by
$$
\delta A_i={\bar\partial}k_i + \sum_{\ell=0}^\infty\sum_{j=0}^{i+2\ell}
f^{j, i-j+2\ell}_{2\ell}(\partial_A,
\partial_k) A_j k_{i-j+2\ell}, \eqno(57)
$$
where $\partial_A$ and $\partial_k$ are the $\partial$ derivatives acting
on $A$ and $k$, respectively.  The term $\hat \delta A_i$ is a co-adjoint
transformation of $A_i$, while the $V^i$ transforms in the adjoint
representation of $W_\infty$; i.e. $\int\Big(\hat\delta A_i V^i+ A_i
\delta V^i\Big)=0$. This model of a complex fermion coupled to $W_\infty$
gauge fields
is in fact intimately related to the model with a single scalar coupled to
$w_\infty$ gauge fields we also just discussed, as will be shown presently.

\bigskip
\bigskip
\noindent
{\twelverm 5. QUANTIZATION DEFORMS $w_\infty$ TO $W_\infty$}
\medskip

The programme of quantization in the case of $W_\infty$ again follows
closely that of Virasoro. Consider an action with local $W_\infty$
symmetry on a two-dimensional ``world-sheet,'' which is a functional of
$W_\infty$ gauge fields denoted collectively by $\cal A$ and generic
matter fields $\Phi$. In the path-integral formalism, the partition
function can be formally written as the following:
$$
{\cal Z}=\int [D{\cal A}][D\Phi] e^{{1\over \pi}\int{\cal L}({\cal A},\Phi)}.
\eqno(58)
$$
One can now proceed in two stages. Firstly one integrates out the
matter fields to arrive at an effective action of the gauge fields
$\cal A$, defined by
$$
\Gamma[{\cal A}]\equiv {\hbox{log}}\int[D\Phi]e^{{1\over \pi}\int{\cal L}}.
\eqno(59)
$$
The second stage consists of quantizing the gauge fields in the
effective action $\Gamma$ that results from integrating out the matter
fields in
the first stage.

The main issue that will be discussed is the quantum consistency of
$W_\infty$ gravity and the $W_\infty$ string. We shall investigate this
issue in
the model of $w_\infty$ gravity coupled with a single scalar that we
discussed above in the classical picture, and demonstrate perturbatively that
the process of quantizing the matter field $\varphi$ coupled to the gauge
fields of
$w_\infty$ gravity ``renormalizes'' the theory to be $W_\infty$ gravity
in order to maintain quantum consistency [34]. This is the first indication
that
the theory of $w_\infty$ makes sense only at the classical level.
This statement will
be strengthened later when quantum $w_\infty$ gravity (coupled to
some matter system, in which case its quantization does not lead to
inconsistency)
is shown to have little dynamics [35].

Quantum consistency requires that, after the completion of the first
stage of quantization, the effective action $\Gamma[{\cal A}]$ be invariant
under gauge transformations on ${\cal A}$. In general this is often not
the case;
an anomaly may occur. For a symmetry realized non-linearly in matter
fields such as the one of $w_\infty$ realized in a single scalar
$\varphi$ given in (45), there are two types of anomalies. The first
type is called universal anomaly, which is only dependent on
the gauge fields themselves. The second is called matter-dependent
anomaly, which explicitly depends on the matter fields that are
integrated out. It is necessary to remove the matter-dependent
anomaly at once, because it is illogical for the effective action
to be dependent on matter fields that are supposed to have been
integrated out in obtaining the effective action. To achieve
this, one adds finite local counter-terms order by order in loops
to the classical Lagrangian. The addition of these terms changes
the couplings between the matter field and the gauge fields,
thus modifying currents of the symmetry in such a way that $w_\infty$
deforms into $W_\infty$. Consequentially the transformation
laws for the gauge fields are modified in response to
the change of the symmetry.

To illustrate precisely how this procedure is implemented, we shall
go over the discussion on the following 1-loop diagrams that give
rise to a matter-dependent anomaly, whose removal deforms the symmetry
structure of the theory.

\bigskip
\boxit{\vskip1.51in
$$
\vbox{\hrule height 0pt depth 0pt width 3.60in \special{picture fig1}}
$$
\hskip3.1in Fig.\ 1\hskip3.1in\smallskip}
\bigskip

    The first diagram that can generate matter-dependent anomalies in the
$w_\infty$ algebra is given in Fig.\ 1.
Its contribution to the effective action is
$$
\eqalign{
\Gamma_{01\varphi}&={1\over\pi^2}\int
d^2z d^2w A_0(z)A_1(w){1\over(z-w)^4}\partial\varphi(w)\cr
&=-{1\over6\pi}\int
d^2zd^2wA_0(z)A_1(w){\partial_z^3\over\partial_{\bar
z}}\delta^2(z-w)\partial\varphi(w)\cr
&=-{1\over 6\pi}\int
d^2z\left({\partial^3\over\bar\partial}A_0(z)\right)A_1(z)
\partial\varphi(z).\cr}\eqno(60)
$$
Under the leading order inhomogeneous terms in the gauge transformations
(2.3) ($\delta A_0=\bar\partial k_0+\cdots$, $\delta A_1=\bar\partial
k_1+\cdots$) the
anomalous variation of $\Gamma_{01\varphi}$ is
$$
\delta\Gamma_{01\varphi} = -{1\over6\pi}\int
d^2z(A_1\partial^3k_0-k_1\partial^3A_0)\partial\varphi.\eqno(61)
$$
Note that in the derivation of this result, one may drop terms proportional
to the $\varphi$ field equation, since these cancel in the quantum Ward
identity [34] against terms involving operator insertions of the $\varphi$
transformations into the relevant one-loop diagrams.

     The anomalous variation (61) can be cancelled by adding the finite
local counterterms $L_{1/2}+L_1$, given by
$$
\eqalignno{
L_{1/2}&= \ft12\Big(A_0 \partial^2\varphi + A_1
\partial\varphi\, \partial^2\varphi\Big),&(62)\cr
L_1&=\ft1{12}  A_1 \partial^3\varphi,&(63)\cr}
$$
and by simultaneously correcting the $\varphi$ transformation (42) by the
extra terms $\delta_{1/2}\varphi +\delta_1\varphi$ given by
$$
\eqalignno{
\delta_{1/2}\varphi&=-\ft12\Big( \partial k_0 + \partial k_1
\, \partial\varphi\Big),&(64)\cr
\delta_1\varphi&=\ft1{12} \partial^2 k_1.&(65)\cr}
$$
One can check that $\delta_{1/2}L_0+\delta_0L_{1/2}=0$, so up to 1-loop
order, the
remaining anomaly-cancelling terms as desired read
$$
\delta_0 L_1 + \delta_{1/2} L_{1/2} +\delta_1 L_0.
\eqno(66)
$$
These variations cancel the anomalies in (61)  completely.

     The occurrence of the counterterms (62) and (63) implies that the
original spin-2 and spin-3 currents of the form (45) have received
corrections, so that they now take the form
$$
\eqalignno{
V^0&=\ft12 (\partial\varphi)^2 + \ft12\partial^2\varphi,
&(67)\cr
V^1&=\ft13(\partial\varphi)^3 +\ft12 \partial\varphi\,
\partial^2\varphi +\ft1{12} \partial^3\varphi.&(68)\cr}
$$
The transformation rules for the matter field $\varphi$, including the
corrections (64) and (65), are precisely those that follow from the
standard expression (38)
with $V^i$ replaced by the expressions given for spin-2 and spin-3 by
(67) and (68).

     One can in principle proceed, by looking at higher-order diagrams with
higher-spin external gauge fields, to determine the appropriate
modifications to all the higher-spin currents that are needed in order to
remove matter-dependent anomalies.  At the same time, the transformation
rules for the $\varphi$ field will require higher-spin modifications too.
As in the sample diagram studied above, the modifications to the
$\varphi$ variation will be precisely those that follow by substituting the
modified currents into (38).  There are further kinds of
matter-dependent anomalies, of types that are not illustrated by the diagram
in Fig.\ 1, whose cancellation requires that the gauge-transformation rules
(48) should also be modified.  To build up the entire structure of the
modifications to currents and transformation rules by these diagrammatic
methods would clearly be a cumbersome procedure.
We shall just consider one more diagram to illustrate the way in which the
gauge-transformation rules (48) must receive corrections.
\bigskip
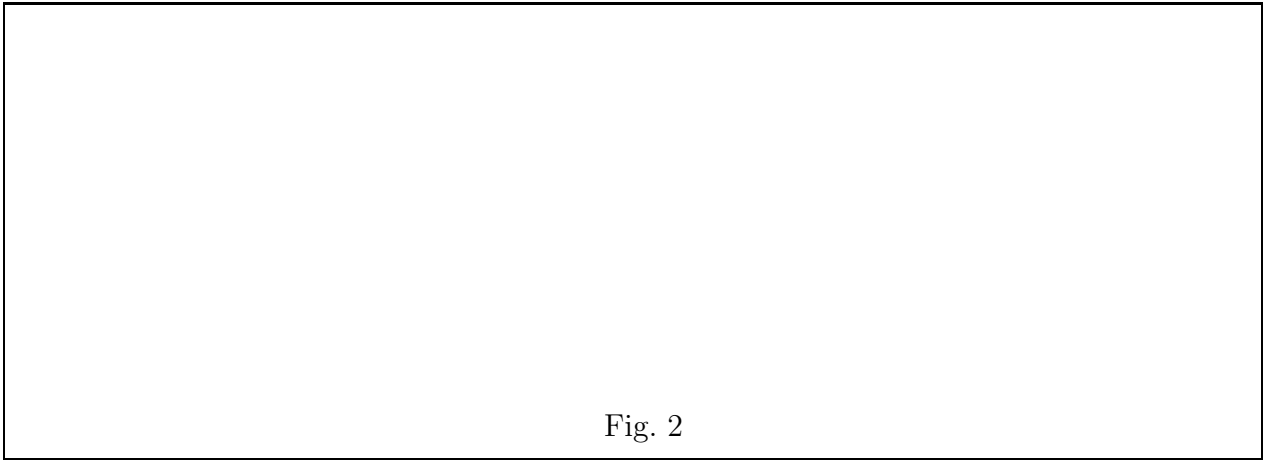
\boxit{\vskip1.56in
$$
\vbox{\hrule height 0pt depth 0pt width 3.60in \special{picture fig2}}
$$
\hskip3.1in Fig.\ 2\hskip3.1in\smallskip}
\bigskip

     The simplest diagram that gives rise to a matter-dependent anomaly
whose removal requires making modifications to the gauge-field
transformation rules is shown in Fig.\ 2.  It produces a contribution
to the effective action given by
$$
\Gamma_{11\varphi\varphi}=-{1\over 6\pi} \int d^2z
A_1(z)\partial\varphi(z)  {\partial^3\over \bar\partial}
\Big(A_1(z)\partial\varphi(z)\Big).  \eqno(69)
$$
This gives rise to an anomalous variation with respect to the leading-order
inhomogeneous term in the $A_1$ variation, {\it i.e.} $\delta A_1={\bar
\partial} k_1+\cdots$:
$$
\delta \Gamma_{11\varphi\varphi}={1\over 3\pi} \int d^2z A_1(z) \partial
\varphi(z) \partial^3\Big( k_1(z) \partial\varphi(z)\Big).\eqno(70)
$$
Cancellation of this anomaly requires, in addition to the modifications
to the spin-2 and spin-3 currents in (67) and (68) and the
modifications (64) and (65) to the $\varphi$ transformation rules, a
correction to the spin-2 gauge transformation rule:
$$
\delta_1 A_0=\ft1{20}  \Big( 2\partial^3 A_1\, k_1 -3\partial^2 A_1 \,
\partial k_1 + 3\partial A_1\, \partial^2 k_1 -2 A_1\, \partial^3 k_1 \Big),
\eqno(71)
$$
and a new counterterm
$$
L_1= A_2\Big( \ft15\partial\varphi\, \partial^3\varphi -\ft1{20}
\big(\partial^2\varphi\big)^2\Big).\eqno(72)
$$
This counterterm implies that the spin-4 current receives quantum
corrections. There are other anomaly diagrams
that give rise to further quantum corrections
to the spin-4 current.
Indeed they turn out to be precisely the spin-4 current of $W_\infty$
realized in terms of a single scalar.

We have now seen how the mechanism for cancelling the matter-dependent
anomalies arising from the diagrams in Fig.\ 1 and Fig.\ 2 leads to
quantum corrections to the currents, and to the matter and gauge-field
transformation rules. These constructions can be carried out to arbitrary
order in principle. For now we shall be content with just these two
examples in drawing the conclusion that quantum consistency
promotes $w_\infty$ to $W_\infty$ in the process of renormalization
(see ref.\ [34] for
more details).

There is in fact a much more elegant way of understanding the
consistent theory of a matter field $\varphi$ coupled to the $W_\infty$
gauge fields, given in [34]. One can ``fermionize'' the field $\varphi$
to obtain a matter system described by a complex fermion $\psi$. The
advantage in taking such a viewpoint is that the $W_\infty$ currents
(obtained from the $w_\infty$ currents in terms of $\varphi$ plus
modifications from renormalization) realized non-linearly in terms of
$\varphi$ convert into bilinears in terms of the fermions, which implies
transformations on $\psi$. In this picture, 1-loop diagrams never
have external $\psi$ legs; $\psi$ only appears in the internal
loops, because of the fact that there exist only three-point couplings
between $\psi$
and the gauge fields $A_i$. Hence there is no matter-dependent anomaly
after the
quantization of $\psi$. Thus as far as the matter-dependent anomaly is
concerned with
regard to quantum consistency, the system of a complex fermion matter
field coupled to
$W_\infty$ gravity makes perfect sense. In fact it is precisely the model
of $W_\infty$
gravity discussed in our previous section.

We now proceed to the second stage in quantizing the gauge fields.
Although we have succeeded in getting rid of the matter-dependent
anomaly, there remains the univeral anomaly, which spoils the
invariance of the effective action under $W_\infty$ gauge
transformations. However the universal anomaly, though undesirable,
is strongly dictated by the structure of $W_\infty$ symmetry (hence
the name ``universal,'' for it is not dependent on any particular matter
system
to realize the symmetry) and has a very simple form. As we shall
show presently, in the light-cone gauge it is given by
$$
\delta_k \Gamma=\sum_{i\ge0}{c_i\over\pi} \int d^2z k_i \partial^{2i+3}
A_i,\eqno(73)
$$
where $c_i$ are related to the total central charge $c_{\rm total}$
of the Virasoro sector of $W_\infty$ (14) .

We shall derive (73) in the operator formalism. Let us define
$$
\big\langle {\cal O}\big\rangle \equiv \int[D\Phi] e^{{1\over\pi}\int
{\cal L}_0}{\cal O},\eqno(74)
$$
where ${\cal L}_0$ is the free Lagrangian of a generic
matter system with its fields denoted by $\Phi$; ${\cal O}$ denotes
a generic operator.
Thus the effective action can be written in this language as follows.
$$
e^{-\Gamma(A_i)}=\big\langle \exp\big(-{1\over\pi}\int \sum_i A_i
V^i\big)\big\rangle. \eqno(75)
$$
Varying (75) with respect to $A_i(z)$, one finds
$$
{\delta\Gamma\over\delta A_i(z)}={1\over\pi}\big\langle
V^i(z)\,\exp\big(-{1\over\pi}\int \sum_j A_j V^j\big)\big\rangle
e^{\Gamma}\eqno(76)
$$
and hence
$$
\bar\partial{\delta\Gamma\over\delta
A_i(z)}={1\over\pi}\big\langle\bar\partial V^i(z)\,\exp\big(-{1\over\pi}\int
\sum_j A_j V^j\big)\big\rangle e^{\Gamma}.\eqno(77)
$$
The occurrence of the $\bar\partial=\partial_{\bar z}$ derivative in (77)
means that the only non-zero contributions will come from $\bar\partial$
acting on singular terms in the operator product expansion of the
operators. Thus, we may calculate
$$\eqalignno{
\bar\partial V^i(z)\exp\big(&-{1\over\pi}\int\sum_j A_j V^j\big)=\bar\partial
V^i(z) \sum_{n\ge0}
{1\over n!}\Big(-{1\over\pi}\int A_j V^j\Big)^n\cr
&=-{1\over\pi}\sum_{n\ge1}{1\over(n-1)!}\int d^2w\big[\bar\partial
V^i(z)V^j(w)\big]A_j(w) \Big(-{1\over\pi}\int A_k V^k\Big)^{n-1},&(78)\cr}
$$
where the brackets around $\bar\partial V^i(z)V^j(w)$ in (78) indicate
that the operator product expansion should be taken just between these
two operators.

Using (33), the operator products in (78) may be evaluated, to give
$$
\eqalign{
&\bar\partial V^i(z)\exp\big(-{1\over\pi}\int\sum_i A_i V^i\big)=\cr
\noss
& {1\over\pi}\int d^2w \partial_{\bar z}
\Big( \sum_{\ell\ge0} f^{ij}_{2\ell}(\partial_z,\partial_w)
{V^{i+j-2\ell}(w)\over z-w}+c_i \partial^{2i+3}_z{1\over z-w} \Big)A_j(w)
\exp\big(-{1\over\pi}\int A_k V^k\big).\cr}\eqno(79)
$$
Since $\partial_{\bar z}{1\over z-w}= \pi \delta^{(2)}(z-w)$, we may perform
the integration in (79).  Thus we find, from (79), that
$$
\bar\partial {\delta\Gamma\over \delta A_i} +\sum_{\ell\ge0}
f^{ij}_{2\ell}(\partial,-\partial_A)\Big( {\delta \Gamma\over \delta
A_{i+j-2\ell}} A_j\Big)=-{c_i\over\pi} \partial^{2i+3} A_i.\eqno(80)
$$
The subscript $A$ on the second derivative argument of $f^{ij}$ indicates
that it should act only on the explicit $A_j$ in the parentheses that
follows it, whilst the first derivative argument of $f^{ij}$ acts on all
terms in the parentheses.  Eq. (80) is the anomalous Ward identity
for $W_\infty$ gravity.

It is easy now to obtain the variation of the effective action
when the gauge fields $A_i$ are transformed according to laws of
$W_\infty$ given in (57).
If we now multiply (80) by the spin-$(i+2)$ transformation
parameters $k_i$ and integrate, we find
$$
\int{\delta\Gamma\over \delta A_i}\Big(\bar \partial k_i +\sum_{\ell\ge0}
\sum_{j=0}^{i+2\ell} f^{i-j+2\ell,j}_{2\ell} (\partial_k,\partial_A)
k_{i-j+2\ell} A_j  \Big) ={c_i\over\pi} \int k_i \partial^{2i+3}
A_i.\eqno(81)
$$
Now we see that the left-hand side of this equation involves
precisely the $W_\infty$ gauge-transformation rule for $A_i$ under $k_j$
given
in (57), and so it can be written simply as $\delta_k \Gamma$,
which is nothing
but the desired result in (73).

Since the universal anomaly spoils the invariance of the effective
action under $W_\infty$ symmetry, it has to be removed in order to
have a fully consistent theory. There are two possibilities. The first
one is that the total central charge vanishes, in which case the
dynamics of the gauge fields decouples from the matter system
involved. In string terminology, this is known to be the critical
case. The other possibility in the light-cone gauge is to restrict
the dynamics of the gauge fields by the following vanishing condition:
$$
\pa^{2i+3} A_i=0.\eqno(82)
$$
This case is often known in string theory as non-critical. Note that
in ordinary string theory, the non-critical situation has the well-known
elegant formulation in the conformal gauge, where the dynamics
of two-dimensional gravity is described by the Liouville field [36].
The Liouville
field contributes to the total stress tensor in such a way that the total
central
charge of both the matter and the Liouville (gravity) sectors amounts
to the critical value.
For the $W_\infty$ string, it is not clear how such an
analogue should be formulated.

It is necessary now to take full account of the central charges.
The contribution from $W_\infty$ matter is simple to account for, {\it
i.e.} the central charge of its Virasoro sector. The other important part
is the
ghost contribution to the total central charge that arises from the need
to avoid
over-counting in the path-integral over the gauge fields, due to the gauge
degrees of freedom. Again this is basically a property of the gauge symmetry
itself,
irrespective of the matter system involved; thus it can be dealt with quite
independently. The most convenient method of removing such a redundancy
arising from gauge degrees of freedom is the well-known
BRST formalism, which is our next topic.

\bigskip
\bigskip
\noindent
{\twelverm 6. THE BRST ANALYSIS OF $W_\infty$}
\medskip

Before plunging into the full BRST analysis of $W_\infty$, which
is rather demanding technically, we shall start with some well-known
results on this issue for the case of the finite-$N$ $W_N$ algebra, and
try to grasp
some ideas about what the result in the limit $N\to\infty$ would be.
It is expected that there might not be a unique limiting procedure,
which is the case for the algebraic structure of finite-$N$ $W_N$
in the large $N$ limit; nonetheless such a strategy may still prove
to be instructive.

For a theory with local $W_N$ gauge symmetry, it is well known that
the ghost contribution to the total anomaly in the Virasoro sector
is given by
$$
c_{\rm gh}=\sum_{s=2}^N c_{\rm gh}(s),\eqno(83)
$$
where
$$
c_{\rm gh}(s)=-2(6s^2-6s+1)\eqno(84)
$$
is the contribution from a pair of ghosts with spins $(1-s,s)$ for
the spin-$s$
gauge fields. The ghosts are necessary in order to remove the over-counting
in the
integration over the spin-$s$ gauge field. Thus, after the
summation, (83) becomes
$$
c_{\rm gh}=-(N-1)(4N^2+4N+2).\eqno(85)
$$
In addition, there will also be ghost contributions to the anomalies in
all the higher-spin sectors. Of course the values of the central-charge
contributions in the various spin sectors are all related to one another,
since there is just one overall central-charge parameter in the $W_N$
algebra.

Na\"\i vely, by setting $N=\infty$ in (85), one would think that the total
ghost contribution in the Virasoro sector of $W_\infty$ would be $c_{\rm
gh}=-\infty$. Such a scenario, though logical, is arguably less
desirable and manageable. A more appealing approach is
to treat the divergent sum (85) over the individual spin-$s$
contributions as a quantity that should be rendered finite by some
regularization procedure [37].  Likewise, the ghost contributions in all the
higher-spin sectors will be given by divergent sums, which can also be
regularized.  The regularization procedures for each spin
must be consistent with one another,
since there is just one overall central-charge parameter
in the $W_\infty$ algebra.  In [38], it was shown that a natural
zeta-function regularization scheme gives the regularized result
$$
c_{\rm gh}=2.\eqno(86)
$$
The trick is to introduce the generalized zeta function
defined through analytic continuation in $s$ of the sum
$$
\zeta(s,a)=\sum_{k\ge0}(k+a)^{-s},\eqno(87)
$$
which converges for $s>1$. Thus (83) can now be written as follows.
$$
c_{\rm gh}=-6\zeta(-2,\ft32)+\ft12\zeta(0,\ft32),\eqno(88)
$$
which gives (86) in the sense of analytical continuation.

A consistent extension of this regularization scheme to all spin sectors
was proposed in [38], where it was shown that it gave consistent results at
least up to the spin-18 level. The fact that such a universal scheme
exists is highly suggestive of an underlying
interpretation and rigorous justification for the regularization
procedure, possibly in terms of a higher-dimensional
theory. Next we shall only review the mechanical procedure of
this regularization scheme, leaving aside the more difficult question of
finding an underlying reason.

The standard prescription for constructing the BRST charge for a Lie
algebra with structure constants $f^{ab}{}_c$ is
$$
Q= c_a T^a -\ft12 f^{ab}{}_c c_a c_b b^c,\eqno(89)
$$
where $c_a$ and $b^a$ are the ghosts (anticommuting for a
bosonic algebra) that satisfy $\{c_a,b^b\}=\delta_a^b$, with
the other anticommutators vanishing.  $Q$ may be written as
$Q=Q_T+ Q_{\rm gh}$, with $Q_T=c_a T^a$ and $Q_{\rm gh}= \ft12
c_a T_{\rm gh}^a$, where $T^a_{\rm gh}=\{Q,b^a\}$ gives a ghostly
realisation of the algebra. The generic index in (89) for the case of
an infinite-dimensional algebra such as Virasoro can be either the $z$
coordinate
of the spin-2 stress tensor or its Fourier mode index. The former is the
BRST
analysis in OPE language, while the latter is somewhat conventional (with
more
indices). Here we shall present the analysis in the Fourier mode convention.
For the $W_\infty$ algebra, given by (8-14), $Q$ in (89) becomes
$$
Q=\alpha_0c_0^0+\sum_{i,m}V^i_mc^m_i-\ft12\sum_{{i,j,\ell,}\atop{m,n}}
g^{ij}_{2\ell}(m,n)\,{:}\,c^{-m}_ic^{-n}_jb^{m+n}_{i+j-2\ell}\,{:}\, ,
\eqno(90)
$$
where $c_i^m$ and $b_i^m$ are the $m$'th Fourier modes of ghosts and
antighosts for spin $i+2$.  In (90) we have allowed for an intercept
$\alpha_0$, expected on general principles due to normal-ordering
ambiguities in the remaining terms.
Since $W_\infty$ is a Lie algebra,
$Q$ is guaranteed to be nilpotent up to central terms.
One finds that
$$
Q^2=\sum_{m>0}c^m_i c^{-m}_j\left(R^{ij}_T(m)+R^{ij}_{\rm gh}(m)\right),
\eqno(91)
$$
where
$$
R^{ij}_T(m)=\delta^{ij}\left(c_i(m)-\alpha_0g^{ii}_{2i}(m,-m)\right),
\eqno(92)
$$
while the contribution from $Q_{\rm gh}^2$ reads
$$
R^{ij}_{\rm gh}(m)=\sum_{r=0}^{(i+j)/2}\sum_{k={\rm max}(0,2r-i)}^\infty
\left\{\sum_{p=1}^m g^{ik}_{2r}(m,-p)\,g^{\,j,k+i-2r}_{i+j-2r}(-m,m-p)
\right\} \eqno(93)
$$
when $i+j$ is even, and zero otherwise (actually,
it turns out that $R^{ij}_{\rm gh}(m)$ vanishes identically for
$i\neq j$, just as $R^{ij}_T(m)$).

At this point one may think that it is straightforward to use the
generalized zeta-function to extract finite answers to these expressions.
However, a little manipulation with them reveals that there exist
many ways of expressing them in terms of the zeta-functions, which
may give different answers, and thus correspond to different
regularization schemes. Thus,
{\it a priori}, one could obtain any result
that one wishes by choosing the regularization scheme appropriately.
However, we know that the generators $V^i_m({\rm gh})$ in $Q_{\rm
gh}=\ft12 c^m_i V^i_m({\rm gh})$ should provide a ghost realization of
the algebra. This means that since the central terms in $W_\infty$ are
uniquely determined up to an overall scale, it follows that all the
central terms in $Q_{\rm gh}^2$ must be regularized in a self-consistent
way in order that their regularized values be consistent with the
Jacobi identities for the algebra.

In [38] it was shown that there is in fact a natural-looking, and easily
specifiable, scheme for regularizing all the central terms in a
consistent manner.  It amounts to first performing a constant shift
$\Delta_{ir}$ of the $k$
parameter in (93), for each value of $r$, in order to make the summand
into an even function of the shifted parameter.
The fact that this can be done is non-trivial.
The shift $\Delta_{ir}$, which turns out to depend upon
$r$ and $i$ (but not upon $j$), is given by $\Delta_{ir}=\ft12(i+3)-r$.
When $r\le i/2$, the summand in (93) will now take the general form
$$
\sum_p\sum_{k\ge 0}A_p(k+\Delta_{ir})^{2p}+\sum_{k\ge0}F(k),\eqno(94)
$$
where $F(k)$ is an absolutely-convergent
sum of simple fractions of the form $1/(k+b)^q$.  The
divergent polynomial sums are then regularized using the generalized
zeta function defined in (87).

For the $W_\infty$ algebra, one finds that that the coefficients
$R^{ij}_{\rm gh}(m)$ in (93) are zero unless $i=j$.  The non-zero
coefficients are precisely of the form $R^{ij}_T(m)$ (with $\alpha_0=0$)
determined by the central terms in the
$W_\infty$ algebra, with a (regularized) central charge $c_{\rm
gh}=2$.  For example, for $i=j=0$ one has
$$
R^{00}_T(m)=\ft1{12}c(m^3-m)-2\alpha_0m,
\eqno(95)
$$
where $c$ is the usual central charge in the matter sector, and
$$
\eqalign{
R^{00}_{\rm gh}(m)
&=\sum_{k\ge0}\Big(-m^3\Big((k+1)^2+(k+1)+\ft16\Big)+\ft16m\Big)\cr
&=\sum_{k\ge0}\Big(-m^3(k+\ft32)^2+\ft1{12}m^3+\ft16m\Big)\cr
&=-m^3\zeta(-2,\ft32)+\big(\ft1{12}m^3+\ft16m\big)\zeta(0,\ft32)\cr
&=\ft16(m^3-m).\cr}\eqno(96)
$$
Thus, requiring that the coefficient of $c_0^{-m}c_0^m$ in $Q^2$
vanish leads to the anomaly-freedom conditions $c=-2$ and $\alpha_0=0$.

One can carry out a similar analysis for $W_{1+\infty}$, in which case the
anomaly-freedom condition is given by [38]
$$
c=0, \qquad \alpha_0=0.\eqno(97)
$$
This means in particular that $W_{1+\infty}$ gravity is consistent by
itself without
coupling to matter.

\bigskip
\bigskip
\noindent
{\twelverm 7. \slinfty  KAC-MOODY SYMMETRY IN $W_\infty$ GRAVITY}
\medskip

Having analyzed the ghost contribution to the total central charge
in the universal anomaly, we are in a position to
obtain a fully consistent quantum theory of $W_\infty$ gravity.
Since the (regularized) ghost contribution $c_{\rm gh}=2$, one needs
a matter system with $c_{\rm matter}=-2$ in order to have $c_{\rm
total}=0$ for the critical $W_\infty$ string. As a matter of fact, the
model with a single scalar $\varphi$ renormalized to realize local
$W_\infty$ symmetry discussed previously is precisely one of
them, since the matter sector has central charge $-2$. In the
corresponding fermion language, the matter system is viewed as
a pair of spin (0,1) $b-c$ systems, which is especially useful, for
example in the bosonization of bosons [39] and two-dimensional
topological gravity [40].

When the central charge of a matter system is not the critical value
$-2$ of the $W_\infty$ string, it is still possible to have a consistent
theory, in which case the $W_\infty$ gauge fields will not decouple. This is
because the gauge fields can be tuned to make up for the difference
between the necessary critical value and the actual value of the central
charge of
matter. In the light-cone gauge, it
amounts to restricting the configuration space of the gauge fields
in such a way that the universal anomaly vanishes, thus giving
rise to a consistent quantum theory. In the case of two-dimensional
gravity coupled some matter system,
such a strategy has proved to be rather fruitful in that
an $SL(2,R)$ Kac-Moody symmetry was discovered by Polyakov [41], from
which the authors of ref.\ [42] were able to extract some
non-perturbative information about the system.

Naturally there arises the question whether such a strategy can also
be applied and generalized to the case of
$W_\infty$ gravity coupled to some $W_\infty$ matter.The most obvious
question is what the analogue of the $SL(2,R)$ Kac-Moody algebra of
two-dimensional quantum
gravity is for quantum $W_\infty$ gravity. In ref.\ [35], it has been
shown that it is the
\slinfty Kac-Moody algebra.

To prove such a statement, let us first recall the
way Polyakov has shown the existence of $SL(2,R)$ Kac-Moody symmetry
in light-cone two-dimensional gravity. Firstly a set of
recursion relations for the spin-2 gauge field was derived from
the anomalous Ward identity of two-dimensional gravity. Secondly since
the spin-2
gauge field is restricted by the anomaly-freedom condition given in (82)
with $i=0$,
there are only three dynamical components $j^a$ ($a=-1,0,1$) in powers
of $z$ as follows.
$$
A_0 =j^{(1)}-2j^{(0)} z +j^{(-1)} z^2.\eqno(98)
$$
One next deduces a set of recursion relations for $j^a(\bar z) $, which
turns out to be precisely that dictated by an $SL(2,R)$ Kac-Moody symmetry
of those $j^a(\bar z)$. Thus one proves the existence of such a symmetry.

This line of logic proceeds essentially unaltered in the case
of $W_\infty$. Firstly we set up our notations. For a generic operator
${\cal O}$ that is a functional of the gauge fields $A_i$ only, its
expectation value is defined by
$$
\ll{\cal O}\rr\equiv\int {\cal D}A e^{-\Gamma}
{\cal O}.\eqno(99)
$$
Here the double-angle brackets are used to be distinguishable from the
single-angle brackets introduced previously in the first stage of
quantization
that correspond to integration over the configuration space of the matter
fields
only. Note that while the single-angle brackets are defined for operators
that can be a functional of both matter fields and the gauge fields,
the double-angle brackets only make sense for operators of the gauge
fields, because matter fields are supposed to have been integrated out
at the second
stage.

Consider the $(n+1)$-point correlation function $\ll A_i(z) A_{j_1}(x_1)
\cdots
A_{j_n}(x_n)\rr$ for the gauge fields $A_i$. Applying the operator
$\pa^{2i+3}_z$ to it, and recalling the $W_\infty$ anomalous Ward identity
given in
(80), we can now write down recursion relations for the correlation
functions of the gauge fields $A_i$. Thus we have
\crampest
$$
\eqalignno{
&-{c_i\over\pi}\partial_z^{2i+3}\ll A_i(z) A_{j_1}(x_1)\cdots
A_{j_n}(x_n)\rr=
\ll {\bar\partial_z}{\delta\Gamma\over\delta
A_i(z)}
A_{j_1}(x_1)\cdots A_{j_n}(x_n)\rr \cr
&+
\sum_{k\ge0}\sum_{\ell=0}^{[(i+k)/2]}f^{ik}_{2\ell}(\partial,-\partial_A)
\ll {\delta\Gamma\over\delta A_{i+k-2\ell}(z)}
A_k(z)A_{j_1}(x_1)\cdots A_{j_n}(x_n)\rr\cr
\cr
&=\sum_{p=1}^n {\bar\partial}\delta(z-x_p)\delta^{i p}
\ll A_{j_1}(x1)\cdots\ax A_{j_p}
(x_p)\cdots A_{j_n}(x_n)\rr\cr
& +\sum_{k\ge0} \sum_{\ell=0}^{[(i+k)/2]}
\sum_{p=0}^n f^{ik}_{2\ell}(\partial_z,-\partial_{A_k}) \delta(z-x_p)
\delta_{i+k-2\ell,j_p}\ll A_k(x_p)A_{j_1}(x_1)\cdots
\ax A_{j_p}(x_p)\cdots A_{j_n}(x_n)\rr.\cr
&&(100)\cr}
$$
In the final term here, the first derivative operator in
$f(\pa,\pa)$, defined in (30), acts only on the $z$ argument of the
delta-function, and the second derivative operator denotes
${\partial\over\partial
x_p}$ acting only on the $A_k(x_p)$ field in the angle brackets.
The $\ax{A}$
indicates that $A$ be taken out of the correlator. The function
$\delta(z-x_p)$
denotes a two-dimensional delta function. The derivation of the
second line in (100)
from the first makes use of the identity
$$
\ll {\delta\Gamma\over\delta A_i}{\cal O}\rr =
\ll{\delta{\cal O}\over\delta A_i}\rr,\eqno(101)
$$
for arbitrary ${\cal O}$, which can be proved by using the definition (99),
and performing a functional integration by parts.

Using the identity
$$
\partial^{2i+3}{(z-x_p)\over (\bar z-{\bar x}_p)} = \pi (2i+2)!
\delta(z-x_p),\eqno(102)
$$
we may now integrate (100) to obtain
$$
\eqalign{
\ll A_i(z)A_{j_1}(x_1)\cdots& A_{j_n}(x_n)\rr
=\cr
 &\sum_{p=1}^n c_{j_p} (2j_p+2)!
\delta^{i j_p}{(z-x_p)^{2i+2}\over (\bar z-{\bar x}_p)^2} \ll
A_{j_1}(x_1)\cdots \ax A_{j_p}(x_p)\cdots A_{j_n}(x_n)\rr\cr
&-\sum_{k\ge 0} \sum_{\ell=0}^{[(i+k)/2]} \sum_{p=1}^n {c_{j_p}(2j_p+2)!
\over c_k
(2k+2)!}\delta_{i+k-2\ell,j_p} f^{ik}_{2\ell}(\partial_z,
-\partial_{A_k})\cr
&\times {(z-x_p)^{2i+2}\over (\bar z-{\bar x}_p)} \ll A_k(x_p)
A_{j_1}(x_1)\cdots \ax A_{j_p}(x_p)\cdots A_{j_n}(x_n)\rr.
\cr}\eqno(103)
$$
We have also,
for convenience, rescaled the gauge fields $A_i$ according to
$$
A_i\rightarrow {1\over c_i(2i+2)!}A_i.\eqno(104)
$$

This recursion relation may be used to calculate arbitrary $N$-point
correlation functions for the gauge fields.  For example, the two-point
function turns out to be
$$
\ll A_i(x,\bar x)A_j(y,\bar y)\rr= c_i(2i+2)!
\delta^{ij}{(x-y)^{2i+2}\over (\bar x-\bar y)^2}.\eqno(105)
$$
Substituting this back into (103), we find that the three-point function is
given by
$$
\eqalign{
\ll A_i(x,\bar x) A_j(y,\bar y) A_k(z,\bar z)\rr&=
c_{j}(2j+2)!f^{ik}_{i+k-j} (\partial_x,\partial_z)
{(y-z)^{2k+2}(x-y)^{2i+2}\over (\bar y-\bar z)^2 (\bar x-\bar y)} \cr
&+c_k(2k+2)! f^{ij}_{i+j-k}(\partial_x,\partial_y) {(z-y)^{2j+2}
(x-z)^{2i+2} \over (\bar z-\bar y)^2 (\bar x-\bar z)}.\cr}\eqno(106)
$$
Using (14), (30) and (31), we find the following expression for the
three-point
function:
$$
\ll A_i(x,\bar x) A_j(y,\bar y) A_k(z, \bar z)\rr
=N_{ijk}
{(x-y)^{i+j-k+1}(y-z)^{j+k-i+1} (z-x)^{k+i-j+1}\over (\bar x-\bar y)(\bar
y-\bar z)(\bar z-\bar x)},\eqno(107)
$$
where $N_{ijk}$ is defined by
$$
N_{ijk}\equiv {(2i+2)!(2j+2)!(2k+2)!\over
(i+j-k+1)!(j+k-i+1)!(k+i-j+1)!}P_{ijk}.\eqno(108)
$$
Here, $P_{ijk}$ is given by
$$
P_{ijk}=\ft12 c_k \phi^{ij}_{i+j-k}.\eqno(109)
$$
$P_{ijk}$ is manifestly symmetric in $i$ and $j$.  Although it is not
manifest, it is in fact totally symmetric in $i$, $j$ and $k$, by virtue
of the identity
$$
c_j\phi^{ik}_{i+k-j}=c_k\phi^{ij}_{i+j-k}.\eqno(110)
$$
Thus we may rewrite $P_{ijk}$ in the manifestly-symmetric form
$$
P_{ijk}=\ft16\Big(c_k \phi^{ij}_{i+j-k} +c_j \phi^{ik}_{i+k-j} +
c_i \phi^{jk}_{j+k-1}\Big).\eqno(111)
$$
The three-point function in (107) is in agreement with the general structure
of three-point function for conformal fields $A_i$, $A_j$ and $A_k$ with
conformal dimensions $(-1-i,1)$, $(-1-j,1)$ and $(-1-k,1)$
respectively [1,2,3].

The entire discussion that we have given above for $W_\infty$ gravity may
be repeated for the case of $W_{1+\infty}$ gravity. $W_{1+\infty}$ is an
algebra of similar type to $W_\infty$ with an additional spin-1 current.
The details of its structure constants, and central terms are given in
section 2. $W_{1+\infty}$ gravity can be obtained straightforwardly
by gauging an additional spin-1 current to $W_\infty$ gravity.
For the case of $W_{1+\infty}$ gravity, we shall again use the
tilded notations introduced in section 2, and the formulae that we have
derived for recursion relations and correlation functions for
for $W_\infty$ gravity hold {\it mutatis mutandis} for $W_{1+\infty}$
gravity, except now the index $i$ is allowed to take -1.

The anomaly-freedom condition given in (82) allows us to
expand the gauge fields $A_i(z, \bar z)$ of
$W_\infty$ gravity as follows:
$$
A_i(z,\bar z)=\sum_{a=-i-1}^{i+1} (-1)^{i+1+a} {2i+2\choose i+1+a}
J^i_a(\bar z) z^{i+1+a},\eqno(112)
$$
where $J^i_a(\bar z)$ are dynamic fields of arbitrary functions of
$\bar z$. Substituting
this into the two-point function (105) and three-point function (107)
for the gauge fields $A$, we obtain the two-point and three-point
functions for the ``expansion coefficients'' $J^i_a(\bar z)$. For
the two-point function, one finds
$$
\ll J^i_a(\bar x)J^j_b(\bar y)\rr= {K^{ij}_{ab}
\over (\bar x-\bar y)^2},\eqno(113)
$$
where the $K^{ij}_{ab}$ are given by
$$
K^{ij}_{ab}= (-1)^{i+1+a} c_i(i+1+a)!(i+1-a)!\delta^{ij}\delta_{a+b,0}.
\eqno(114)
$$
After some algebra, one finds that the three-point function for $J^i_m$
can be written as
$$
\ll J^i_a(\bar x)J^j_b(\bar y)J^k_c(\bar z)\rr =
{Q^{ijk}_{abc}\over(\bar x-\bar y)(\bar y-\bar z)(\bar z-\bar x)},
\eqno(115)
$$
where the coefficients $Q^{ijk}_{abc}$ are given by
$$
\eqalign{
&Q^{ijk}_{abc} = \delta_{a+b+c,0} \cr
\times &\sum_{d=0}^{k+i-j+1}{(i+1+a)!(i+1-a)!(j+1+b)!(j+1-b)!
(k+1+c)!(k+1-c)!P_{ijk}(-)^{j+1-a+c+d}\over (j-k-a+d)!(i+1+a-d)!
(j-i+c+d)! (k+1-c-d)! (k+i-j+1-d)!d!}\cr}\eqno(116)
$$

As discussed in the previous section, we may repeat the above analysis
for the case of $W_{1+\infty}$ gravity; the expressions above will then
be replaced by analogous tilded expressions.

Our goal now is to compare the two-point and three-point functions
(105)-(107)
for $J^i_a$ with those dictated by an \slinfty Kac-Moody symmetry.
To do this, we start by setting up our notations
for \slinfty. In the literature, a certain class of \slinfty
has been discussed rather extensively [19,13],
where \slinfty is viewed as the tensor algebras of $SL(2,R)$. As
discussed earlier, one of them is related to the $W_\infty$ algebra in
much the same way as $SL(2,R)$ is a subalgebra of the Virasoro algebra.
In this context, \slinfty was termed the ``wedge'' algebra of
$W_\infty$. Owing to this intimate connection between \slinfty and
$W_\infty$, many notations ({\it e.g.} index structure) for \slinfty
appear to be $W_\infty$-like. So special care is necessary to
tell them apart. Particularly it is important to emphasize that, just
as the $SL(2,R)$ subalgebra of Virasoro does not directly bear
any relevance to the $SL(2,R)$ Kac-Moody symmetry
in the case of gravity, so the ``wedge'' algebra \slinfty
of $W_\infty$ does not {\it per se} imply the existence of an \slinfty
Kac-Moody symmetry for $W_\infty$ gravity.  Nonetheless, as we shall see,
such a remarkable symmetry does exist.

We shall give the structure
constants for the 1-parameter family of $GL(\infty,R)$ with generators
$V^i_m$, for which the $m$ index is restricted in the range given by
$$
-i-1\le m \le i+1,\eqno(117)
$$
and $i$ taking values $\ge -1$. This is the 1-parameter family of tensor
algebras
discussed in Sec. 2. Thus the algebras have the following commutation
relations [13]
$$
[V^i_m,V^j_n]=\sum_{\ell\ge0} g^{ij}_{2\ell}(m,n;s)V^{i+j-2\ell}_{m+n},
\eqno(118)
$$
where
$$
g^{ij}_\ell(m,n;s)={\phi^{ij}_\ell(s)\over 2(\ell+1)!} N^{ij}_\ell(m,n),
\eqno(119)
$$
and
$$
\phi^{ij}_\ell(s)=\FF43{-\ft12-2s\. \ft32+2s\. -\ft12\ell -\ft12\. -\ft12\ell}
{-i-\ft12\. -j-\ft12\. i+j-\ell+\ft52}1,\eqno(120)
$$
and $N^{ij}_\ell(m,n)$ is the same as that given in (11) (except the
restriction on
the index $m$ (117)). Note that these quantities are defined here for
all integer
values of the subscript argument, although only those with {\it even}
values occur
in (118). Odd values for the subscript argument will play an important
role presently. Note also that the quantities
$\phi^{ij}_{2\ell}$ and ${\tilde\phi}^{ij}_{2\ell}$ introduced in (13) and
(23) correspond to $\phi^{ij}_{2\ell}(s)$ with $s=0$ and $s=-\ft12$
respectively, which implies that these two tensor algebras are the wedge
algebras of
$W_\infty$ and $W_{1+\infty}$ respectively. In fact, as discussed in [13],
$W_\infty$
and $W_{1+\infty}$ can be viewed as the analytic extensions
``beyond the wedge'' of the
$s=0$ and $s=-\ft12$ $GL(\infty,R)$ algebras. It is precisely these
two tensor
algebras, which we call \slinfty and $GL(\infty,R)$ respectively,
whose corresponding
Kac-Moody algebras emerge in $W_\infty$ gravity and $W_{1+\infty}$ gravity.

We are now ready to discuss the correlation functions and recursion
relations for this family of $GL(\infty,R)$ Kac-Moody algebras,
to show how they are related to our results
for $W$ gravity correlation functions. For reasons that
will become clear shortly, it is convenient at this stage to discuss first
the case for $GL(\infty,R)$ with $s=-\ft12$, and its relation to
$W_{1+\infty}$
gravity.

For an arbitrary algebra $G$, one can write down recursion relations for
Kac-Moody currents $j^A(\bar z)$ [43]:
$$
\eqalign{
\ll j^A(\bar z) j^{B_1}({\bar x}_1) \cdots j^{B_n}({\bar x}_n)
\rr &=
-{K\over 2} \sum_p {\eta^{A B_p}\over (\bar z-{\bar x}_p)^2}
\ll j^{B_1}({\bar x}_1)\cdots
\ax j^{B_p}({\bar x}_p) \cdots j^{B_n}({\bar x}_n) \rr
\cr &+\sum_p
{f^{AB_p}{}_{C_p}\over(\bar z-{\bar x}_p)}\ll
j^{C_p}({\bar x}_p)j^{B_1}({\bar x}_1)\cdots
\ax j^{B_p}({\bar x}_p)\cdots j^{B_n}({\bar x}_n)
\rr.\cr}\eqno(121)
$$
Here, $\eta^{AB}$ is the Cartan-Killing metric, $f^{AB}{}_C$
are the structure constants, and $A,B,\ldots$ are adjoint
indices for $G$. Normally, for a finite-dimensional algebra,
one defines the Cartan-Killing metric by means of the trace of
generators $T^A$, {\it i.e.}\ as $\eta^{AB}\equiv -2{\rm Tr}(T^A\, T^B)$.
For an infinite-dimensional algebra such as $GL(\infty,R)$, this can be
problematical.  We can, however, define a Cartan-Killing metric,
{\it i.e.}\ a symmetric 2-index invariant tensor, in the following way.

It was shown in [13,14] that the $W_\infty$ and $W_{1+\infty}$ algebras,
and hence in
particular their wedge subalgebras, can be viewed as being derived from
some
corresponding associative-product algebras. This is in fact true for
the whole family
of the $GL(\infty,R)$ algebras under discussion [13,14]. In the case of
$W_\infty$,
these multiplications are called the ``lone-star product.''
For the family of the
$GL(\infty,R)$ algebras, the operation is basically tensor product,
modulo certain
ideal. Explicitly they take the form for $GL(\infty,R)$
$$
V^i_m\star V^j_n=\ft12\sum_{\ell\ge -1}
g^{ij}_\ell(m,n;s) V^{i+j-\ell}_{m+n}. \eqno(122)
$$
The structure constants are antisymmetric under the interchange of $(im)$
with $(jn)$ when $\ell$ is even, and symmetric when $\ell$ is odd.
Thus the commutator in (118) may be written as
$$
[V^i_m,V^j_n]=V^i_m\star V^j_n - V^j_n\star V^i_m.\eqno(123)
$$

The lone-star product for the $W_{1+\infty}$ algebra contains the spin-1
generators $V^{-1}_m$.  It turns out that the generator $V^{-1}_0$
commutes with all other generators in the algebra; thus it may be
viewed as the identity operator in the algebra [14]. This enables us to
define an invariant 2-index symmetric tensor, sidestepping the problem
mentioned previously of how to attach a meaning to the operation of
taking the trace of products of generators. In other words the
lone-star algebra provides us with a rule for extracting the singlet
part in the symmetric product of two generators; this is precisely
the function played by the trace operation in the usual definition of
a Cartan-Killing metric.  Thus we define the Cartan-Killing metric as
$$
\eta^{ij}_{mn}(s)\equiv g^{ij}_{i+j+1}(m,n;s).\eqno(124)
$$
For $W_{1+\infty}$, the parameter $s$ takes the value $-\ft12$.
In this case we write
$$
{\tilde \eta}^{ij}_{mn}\equiv g^{ij}_{i+j+1}(m,n;-\ft12),\eqno(125)
$$
in accordance with our previous notation. For $GL(\infty,R)$, $s$ can
take generic values.

Having defined an invariant Cartan-Killing metric for $GL(\infty,R)$, we
are now in a position to compare the correlation functions for the
$GL(\infty,R)$ Kac-Moody algebra with those that we obtained from
$W_{1+\infty}$ gravity. From (121), the two-point function for
$GL(\infty,R)$
Kac-Moody currents $J^a_m(\bar z)$ ($i\ge-1$; $-i-1\le a \le i+1$) is
given by
$$
\ll J^i_a(\bar x)J^j_b(\bar y)\rr= -{K\over 2} {{\tilde\eta}
^{ij}_{ab}\over (\bar x-\bar y)^2}.\eqno(126)
$$
This should be compared with
our expression (113) obtained from $W_{1+\infty}$
gravity.  Equivalence of the two expressions would require that
${\tilde\eta}^{ij}_{ab}$ and ${\tilde K}^{ij}_{ab}$ should be related by
$$
-{K\over2}{\tilde\eta}^{ij}_{ab}= {\tilde K}^{ij}_{ab},\eqno(127)
$$
for some value of the constant $K$. (Recall that ${\tilde K}^{ij}_{ab}$
is the analogue of (113) for the
case of $W_{1+\infty}$ gravity; {\it i.e.}\ with $c_i$ replaced
by ${\tilde c}_i$, given by (24). We are using ${\tilde\eta}^{ij}_{ab}$
for the Cartan-Killing metric, since $s=-1/2$ for the
$GL(\infty,R)$ is the wedge subalgebra of $W_{1+\infty}$.) One can
verify that
(127) does indeed hold, with $K$ given by$K=-\ft14$.

     For the three-point function, one finds from (121) that the result for
$GL(\infty,R)$ is given by
$$
\ll J^i_a(\bar x)J^j_b(\bar y)J^k_c(\bar z)\rr={{\tilde f}^{ijk}_
{abc} \over (\bar x-\bar y)(\bar y-\bar z)(\bar z-\bar x)},\eqno(128)
$$
where ${\tilde f}^{ijk}_{abc}$ denote the structure constants of the
$GL(\infty,R)$ wedge subalgebra of $W_{1+\infty}$ with all three indices
``upstairs.''  (We use the location of the ``spin'' indices $i,j,k$
to define the notion of upstairs
and downstairs.)  The ``two up, one down'' structure constants,
which one reads off directly from the commutation relations (118), are
defined in general by
$$
[V^i_m,V^j_n]=f^{ijp}_{mnk} \,V^k_p\eqno(129)
$$
(summed over ``spin'' index $k$ and Fourier-mode index $p$), and so
$$
{\tilde f}^{ijp}_{mnk}=g^{ij}_{i+j-k}(m,n;-\ft12)\delta_{m+n,p}.\eqno(130)
$$
The downstairs index can then be raised using the Cartan-Killing metric
defined by (124) (with $s=-\ft12$), to give
$$
{\tilde f}^{ijk}_{mnp}=
{\tilde \eta}^{k\ell}_{pq}\,\,{\tilde f}^{ijq}_{mn\ell}=
\sum_\ell
g^{k\ell}_{k+\ell+1}(p,m+n;-\ft12)g^{ij}_{i+j-\ell}(m,n;-\ft12).\eqno(130)
$$
Comparing with our expression (115) for $W_{1+\infty}$ gravity
(with $Q^{ijk}_
{abc}$ replaced by the appropriate tilded version, as described in the
previous section), we find that indeed
$$
{\tilde f}^{ijk}_{abc}=8{\tilde Q}^{ijk}_{abc}.\eqno(131)
$$
Thus the three-point functions derived on the one hand from
$W_{1+\infty}$ gravity, and on the other hand from
$GL(\infty,R)$ Kac-Moody currents (with the parameter $s$
chosen to have the value $s=-\ft12$ appropriate to the $GL(\infty,R)$
wedge subalgebra of $W_{1+\infty}$) are in agreement.  As we shall discuss
at the end of this section, one can also establish the equivalence
of the general recursion relations for correlation functions for
$W_{1+\infty}$ gravity and $GL(\infty,R)$ Kac-Moody currents. Thus we
have established that $W_{1+\infty}$ gravity has a hidden $GL(\infty,R)$
Kac-Moody symmetry, generalizing the $SL(2,R)$ symmetry of two-dimensional
gravity
found by Polyakov.

The situation is a little more subtle for the case of $W_\infty$ gravity.
The reason for this is that for the hidden Kac-Moody algebra in this case,
the
corresponding tensor algebra $SL(\infty,R)$ turns out to be the wedge
subalgebra of
$W_\infty$ generated by $V^i_m$ with $i\ge 0$, which does not contain
the spin-1
generator at the apex. Since there is no spin-1 current in the algebra,
our procedure for defining an invariant Cartan-Killing metric
breaks down in this case.  (The expression (124) vanishes inside the wedge
if $s$ is chosen to have the special value $s=0$.)  However, since the
expression (124) is non-degenerate for all other values of $s$, we may
approach
$s=0$ via a limiting procedure, in which we first rescale the generators,
$V^i_m\to s V^i_m$, before sending $s$ to zero.  Although other structure
constants in the lone-star product will now diverge, the relevant ones
relating $V^i_m$ and $V^j_n$ to $V^{-1}_0$ will now be finite, and this
is sufficient for the purpose of obtaining an $SL(\infty,R)$-invariant
symmetric 2-index tensor. Thus we may define the Cartan-Killing metric
for \slinfty as
$$
\eta^{ij}_{mn}\equiv {d\over ds} g^{ij}_{i+j+1}(m,n;s)\Big|_{s=0}.\eqno(133)
$$
This can be recast in the form
$$
\eta^{ij}_{mn}={\Psi^{ij}_{i+j+1}\over 2(i+j+2)!}N^{ij}_{i+j+1}(m,n),
\eqno(134)
$$
where
$$
\Psi^{ij}_\ell=-\sum_{k=0}^{[(\ell+1)/2]}{4k\over (4k^2-1)} {(-\ft12)_k
(\ft32)_k  (-\ft12\ell -\ft12)_k
(-\ft12\ell)_k\over k!(-i-\ft12)_k (-j-\ft12)_k
(i+j-\ell+\ft52)_k}.\eqno(135)
$$

One can now verify that the two-point function (113) derived from $W_\infty$
gravity,
and the two-point function (126) for $SL(\infty,R)$ Kac-Moody currents, with
$\eta^{ij}_{ab}$ given by (134) and (135), in fact coincide; specifically,
we find that
$\eta^{ij}_{ab}=8 K^{ij}_{ab}$. Similarly, the three-point functions
coincide, where
the downstairs index on the \slinfty structure constants is raised using our
definition (133) for the Cartan-Killing metric; we find that
$f^{ijk}_{abc}=8 Q^{ijk}_{abc}$. Again, as we shall discuss presently,
the equivalence of the
$W_\infty$ gravity and \slinfty Kac-Moody recursion
relations can also be established in general. Thus we see that $W_\infty$
gravity has an underlying \slinfty Kac-Moody symmetry.

In order to demonstrate that the recursion relations for correlation
functions of the $J^i_m(\bar z)$ that follow by substituting (112) into the
$W$ gravity recursion relations are the same as the Kac-Moody recursion
relations (121), only a little more algebra than we have already carried
out is required. The following identity,
$$
\eqalign{
&f^{ik}_{2\ell}(\partial_z,-\partial_{A_k}) (z-x_p)^{2i+2} A_k(x_p)=\cr
&\sum_{c=-k-1}^{k+1}\sum_{d=-i-1}^{i+1}(-)^{i+k+c+d+1} {2i+2 \choose i+1+d}
{2k+2\choose k+1+c} J^k_c({\bar x}_p) x_p^{i+k+c+d+1-2\ell} z^{i+1-d} g^{
ik}_{2\ell}(d,c;s)\cr}\eqno(136)
$$
can easily be established, where the $f^{ik}_{2\ell}$ quantities on the
left-hand side are untilded when $s$ is chosen to be 0 ( {\it i.e.}\ for
$W_\infty$ gravity), and tilded when $s$ is chosen to be $-\ft12$
( {\it i.e.}\
for $W_{1+\infty}$ gravity).  Using this result in (136), the proof of the
equivalence of the Kac-Moody and $W$ gravity recursion relations follows
after some straightforward combinatoric manipulations.

To recapitulate, we have shown that there exists an \slinfty Kac-Moody
symmetry
in quantum $W_\infty$ gravity, in close parallel to the existence of
$SL(2,R)$
Kac-Moody symmetry in two-dimensional quantum gravity. However, there
is one subtle point
in the case of quantum $W_\infty$ gravity that two-dimensional quantum
gravity does not share. {\it A priori}, there are many inequivalent \slinfty
algebras parametrized by $s$, which can be associated with quantum $W_\infty$
gravity, while there is a unique $SL(2,R)$ Kac-Moody algebra.
It is natural as well as remarkable that the underlying Kac-Moody symmetry
for quantum $W_\infty$ picks a specific \slinfty algebra that turns out to be
exactly the wedge algebra of $W_\infty$ itself. This point begs for some
deeper understanding on the interplay between conformal algebras and
current algebras in two-dimensional gauge theories, which has been discussed
generously in literature.

To close we note that, since central extensions are not allowed for the
generators
of $w_\infty$ except the Virasoro sector, one can quickly deduce from
the recursion relations
of $w_\infty$ gravity that all correlators vanish. This indicates that
the dynamics of
$w_\infty$ is very simple if not trivial. After all, $w_\infty$ tends
to be inconsistent at
quantum level and become $W_\infty$ gravity upon renormalization, as
illustrated in the
previous section.

\bigskip
\bigskip
\noindent
{\twelverm 8. SYMMETRY IN THE $c=1$ STRING MODEL}
\medskip

There has been a considerable amount of activity in the study of the
lower-dimensional
string theories in the past two years, which was pioneered in [41,42].
The initial
breakthrough was the discovery of non-perturbative solutions to
two-dimensional quantum
gravity coupled to some matter system [44] obtained by applying the
techniques of matrix
models. This success has led to solutions of various matrix models,
 which often have physical
interpretations as two-dimensional gravity coupled to certain matter system.
More importantly, the non-perturbative information extracted by this somewhat
indirect
means has stimulated a whole range of approaches to formulating and solving
the
problem of two-dimensional gravity coupled to matter. They include
topological
field theory [45,46], continuum Liouville field theory [36,40,47]
and effective
field theory [48].

Soon after the initial breakthrough it was realized  within the framework
of matrix models that much of the non-perturbative
information is encoded in some generalized $KdV$ hierarchy [49]. On
the other hand,
topological field theory re-interprets the solutions of matrix models
and supplies
the mathematical foundation for the solvability of these models [45].
This has led to the
discovery of the so-called Virasoro constraints and $W$ constraints that
dictate the solutions by giving rise to a set of recursion relations for the
physical correlators [46], which suggest some underlying symmetry structure
for
two-dimensional gravity coupled to matter with $c\le1$.

Since the matrix-model approach to two-dimensional quantum gravity coupled
to matter
with the central charge $c$ is limited to be powerful only when $c\le 1$,
it is
then especially interesting to understand the model with $c=1$ in
order to probe the region where $c>1$. Since the models with $c<1$ are shown
to
be described by $W_N$ constraints, it is natural to expect that, as $c\to 1$,
$c=1$ string theory is dictated by $W_\infty$ constraints. There is abundant
but
confusing literature on this point. One interesting success is that various
finite-$N$ $W_N$ constraints can be embedded in $W_{1+\infty}$ constraints in
the context of fermion Grassmannian [28].

In the meantime, continuum Liouville field theory, which was formulated for
qauntum gravity in two dimensional space-time in [36], has successfully
reproduced some of the
results of the matrix models and topological field theory. For example, some
correlation functions are calculated in the context of using the Liouville
field to
describe two-dimensional gravity coupled some conformal matter [47], which
reproduce those
given by the solutions of matrix models and topological field theory. The
advantage of this approach is that it offers an intuitively more physical
picture of
two-dimensional gravity so that many conventional techniques
can be applied. One example is the successful application of BRST
analysis to the
physical states of two-dimensional gravity coupled to conformal matter
with $c\le 1$ [7],
which showed that there exist many new states with non-vanishing ghost
number. In fact,
indications for the existence of these new states first arose in the
calculation of
correlation functions, by both the continuum Liouville method [50] and
matrix-model
analysis [51], where they appear as poles in the correlation functions.
The existence
of these novel states is indicative of some large underlying symmetry.

More recently it has been elegantly shown that there exists a so-called
ground ring
in the space of special physical states of the $c=1$ string theory [10],
on which there
act some large symmetry groups. It turns out that these symmetry groups
are,
roughly, some area-preserving diffeomorphisms and volume-preserving
diffeomorphisms
[10]. Shortly afterwards it was shown in the continuum Liouville theory [8],
by a
different means from that of ref.[10], that indeed there exist symmetry
algebras with
spin-1 vertex operators as their generators, whose
structure constants are identical to that of the area-preserving
diffeomorphism of a
two-dimensional surface.

For the $c=1$ model described by the two-dimensional critical string
with coordinates
$X^\mu=(\phi, X)$ and a linear dilation background
$$
S={1\over 4\pi\alpha^\prime}\int\Big(\partial_a X\partial^a X +
\partial_a \phi
\partial^a \phi-{\sqrt\alpha^\prime}\phi R^{(2)}\Big),\eqno(137)
$$
one has the following vertex operators characterized by labels $J$ and $m$:
$$
\Psi_{J\ m}(z)=\psi_{J\ m} e^{(J-1)\phi},\eqno(138)
$$
where the $\psi_{J\ m}$ are primary fields that form $SU(2)$ multiplets
with $J$
either integer or half-integer, and $m=(-J,-J+1,\ldots,J-1,J)$;
they are constructed
by hitting $:e^{iJX}:$ repeatedly with the $SU(2)$ lowering operator
$H_-(z)$ as
follows.
$$
\psi_{J\ m}(z)\sim \big[H_-(z)\big]^{J-m}:e^{iJX(z)}:\eqno(139)
$$
Here we have introduced the $SU(2)$ generators given by
$$
\eqalign{
H_{\pm}(z)=\oint{du\over 2\pi i}:e^{\pm iJX(u+z)}:\cr
H_3(z)=\oint{du\over 4\pi}\partial X(u+z).\cr}\eqno(140)
$$
Thus the following algebra for the vertex operators has been obtained [8]:
$$
\Psi_{J_1\ m_1}(z)\Psi_{J_2\ m_2}(w)\sim {J_2 m_1- J_1
m_2\over z-w}\Psi_{J_1+J_2-1,m_1+m_2}(w). \eqno(141)
$$

Since the vertex operators are gravitationally dressed to be spin-1 fields,
they form
a current algebra. The group structure constants are extremely simple,
and in fact
identical to that of the $w_\infty$ algebra (21) after a proper shift
in the index $J$. However
there is a crucial difference in that in Eq.(141) the index $m$ is
restricted to be with the
wedge, so to speak, to form multiplets of $SU(2)$, while in Eq.(21)
the Fourier index
of $w_\infty$ generators runs from $-\infty$ to $+\infty$, which
is essential to making
the connection between $w_\infty$ and an area-preserving diffeomorphism,
as shown in
Sec. 2. Thus the vertex operators in Eq.(138) fill up exactly the
wedge algebra of
$w_\infty$, but not the whole $w_\infty$. Since $w_\infty$ is a
contraction of $W_\infty$, its
wedge subalgebra is also a contraction of the wedge subalgebras of
$W_\infty$ or an
$SL(\infty)$ algebra, which we shall call $SL_c(\infty)$. It is
important to realize that
$SL_c(\infty)$ as a subalgebra of $w_\infty$ does not have the
interpretation of an
area-preserving diffeomorphism in the sense discussed in Sec. 2.

The upshot is that the internal group of the current algebra in Eq.(141)
is not
$w_\infty$, but rather $SL_c(\infty)$. In other words the operator algebra
is an
$SL_c(\infty)$ Kac-Moody algebra. The authors of ref.[8] went on to
conjecture that
when a non-zero cosmological constant term $\lambda\int e^{-\phi}$ is
introduced
in the Lagrangian (137), there will be additional terms on the
right-hand side of Eq.(141) so
that the internal group for the operator algebra will be the
wedge algebra of $W_\infty$ or,
more precisely, an $SL(\infty)$ algebra. Since there are a
family of $SL(\infty)$ parametrized
by $s$ given in Sec. 2, and all of which contract to the same
$SL_c(\infty)$, it is not
possible to see which $SL(\infty)$ algebra corresponds to the
internal group of the vertex
operator algebra with non-vanishing cosmological constant.
So far it remains a formidable task
to carry out a direct calculation for the structure constants of
this algebra. Some efforts
have been made in this direction [58].

If this conjecture turns out to be true, there are a few implications.
First of all,
since the $SL(\infty)$ algebras do have an interpretation as area-preserving
diffeomorphisms {\it a l\`a} Hoppe [18,19], the internal group of the
vertex operator algebra
would thus regain a geometrical flavor. Secondly if the internal
$SL(\infty)$ turns out to be
the one that correspond exactly to the wedge subalgebra of $W_\infty$,
these vertex
operators can be plugged into Eq.(112), giving rise to the gauge
fields of $W_\infty$
gauge theory. Thus for each multiplet of $SL(2)$ vertex operators,
one has a
$W_\infty$ gauge field, which carries higher spin on the world-sheet.
This may suggest
that the theory of $W_\infty$ gravity discussed earlier is of
relevance to the $c=1$
string model.

Unfortunately the vertex operators given in Eq.(138) and (139) are not
the whole story;
there are many more BRST invariant operators with ghost number zero
as well as that
with non-vanishing ghost numbers [7,52]. It seems an insurmountable
to carry out an
explicit evaluation on the algebra for the full set of physical operators.

\bigskip
\bigskip
\noindent
{\twelverm 9. SUMMARY}
\medskip

In this paper we have attempted to give a overview of $W_\infty$
theory at both the classical
and quantum levels. We started with realizations of $W_\infty$ and
proceeded to build
$W_\infty$ gravity and a $W_\infty$ string model.  We have included
two specific models, the
first of which is $w_\infty$ gravity coupled to a real scalar, while
the second is $W_\infty$
gravity coupled to a complex fermion. Quantum mechanically the first
model that was discussed
is not consistent and suffers from matter-dependent anomalies, the
removal of which forces the
theory to become a model of $W_\infty$ gravity coupled to a scalar.
Now viewed
from a different standpoint where the scalar is fermionized, the
renormalized model
corresponds to the  second model we discussed for $W_\infty$ gravity
coupled to a
complex fermion that is completely consistent at quantum level,
free from both
matter-dependent and universal anomalies.

In quantizing the guage fields as well as matter fields, we arrived at
the anomalous
$W_\infty$ Ward identities. We next reviewed the BRST analysis of
$W_\infty$ and
showed that the anomaly-freedom condition for the $W_\infty$ string
is that the central
charge of $W_\infty$ matter must be $-2$. We then showed that
there exists an
underlying \slinfty Kac-Moody symmetry in $W_\infty$ gravity.

Finally we made a short excursion into the recent investigation on the
lower-dimensional strings. In particular we looked into the $c=1$ model,
for which a
vertex operator algebra was worked out in ref.[8]. This current algebra
turns out to
be a Kac-Moody algebra with its internal group being a contracted
$SL(\infty,R)$ algebra.
Its structure constants are identical to that of $w_\infty$.

Undoubtedly we have left out a great deal of topics in the theory of
$W_\infty$. Many
of them are not only interesting mathematically but also potentially
relevant to
string theo string theory.
We shall mention a few of them here to conclude our discussion.

Extensions of $W_\infty$ to supersymmetric $W_\infty$ are certainly
very interesting
algebraically. It was shown in ref.[26] that there exists an $N=2$
supersymmetric
$W_\infty$ algebra. From the vantage point of field theory, this
implies that one can
build supersymmetric $W_\infty$ gravity and a $W_\infty$ string [33].
The quantization of
these theories should follow a similar line to our analysis outlined above.
There are
also other extensions of $W_\infty$ [25,53], the gauge theories of
which would also be
interesting.

Since the concept of topological field theory was introduced [54],
it has flourished
in its application to two-dimensional models [45,46]. Analogous to ordinary
two-dimensional gravity, the so-called two-dimensional topological
gravity can also
be generalized to topological $W$ gravity [55]. However,
due to the lack of solid
mathematical foundation for these topological $W$ gravities,
their utility still remains
to be seen.

There is also a considerable amount of interest in understanding finite-$N$
$W_N$ gravity and the $W_N$ string. The non-linearity in the
symmetry of $W_N$ introduces
as much novelty as the difficulty it causes. An overview of
this field can be found in other papers such as ref.[56].

Finally the concept of universal $W$-algebra that
encompasses all finite-$N$ $W_N$ algebras as its truncations
has been pursued
extensively. Although it is not completely clear what the conclusion is,
there
have been some notable developments. In the context of field theory,
it was shown in
[32] that classical $W_N$ gravity can be obtained as a truncation
from the classical
$w_\infty$ gravity. It has also been shown that field theoretic
realization of
$W_\infty$ at $c=-2$ contains realizations of $W_N$ at $c=-2$ and
gives rise to
complete consistent $W_N$ structure constants [57]. However,
these encouraging signs are
far from a proof that the linear $W_\infty$ algebra is the
universal $W$-algebra; there
remains a good possibility that some non-linear generalization of
$W_\infty$, as advocated in
[11], may prove to be the true universal $W$-algebra.

\bigskip
\bigskip
\bigskip
\noindent
{\bf Acknowledgements}
\bigskip
I am indebted to my collaborators for some of my past work reviewed
in this paper. I should
like to gratefully acknowledge the partial support from a World
Laboratory Scholarship.
\bigskip
\vfil\eject

\bigskip
\centerline{\bf REFERENCES}
\bigskip

\item{[1]} A.A. Belavin, A.M. Polyakov and A.B. Zamolodchikov, Nucl. Phys.
{\bf B241} (1984) 333.

\item{[2]} J.L. Cardy, ``Conformal Invariance and
Statistical Mechanics,'' Lecture notes in
Les Houches, Session XLIX, 1988.

\item{[3]} P. Ginsparg, ``Applied Conformal Field Theory,''
Les Houches lecture notes, 1988.

\item{[4]} D. Friedan, Z. Qiu and S. Shenker, Phys. Rev. Lett.
{\bf 52} (1984) 1575.

\item{[5]} A.B. Zamolodchikov, Teo. Mat. Fiz. {\bf 65} (1985) 347.

\item{[6]} V.A. Fateev and S. Lukyanov,  Int. J. Mod.  Phys.
{\bf A3} (1988) 507;
\item{} F.A. Bais, P. Bouwknegt, M. Surridge and K. Schoutens,
Nucl. Phys. {\bf B304}
(1988) 348.

\item{[7]} B. Lian and G. Zuckerman, Phys. Lett. {\bf B254} (1991)
417; Phys. Lett.
{\bf B266} (1991).

\item{[8]} I. Klebanov and A.M. Polyakov, Mod. Phys. Lett.
{\bf A6} (1991) 3273.

\item{[9]} J. Avan and A. Jevicki, Phys. Lett. {\bf B266} (1991) 35;
preprints
BROWN-HET-824, BROWN-HET-839;
\item{} D. Mimic, J. Polchinski and Z. Yang, preprint UTTG-16-91;
\item{} G. Moore and N. Seiberg, preprint RU-91-29, YCTP-P19-91;
\item{} S. Das, A. Dhar, G. Mandal and S. Wadia, preprint IASSNS-HEP-91/52,
IASSNS-HEP-91/72.

\item{[10]} E. Witten, ``Ground Ring of Two-dimensional String Theory,''
preprint
IASSNS-HEP-91/51, 1991.

\item{[11]} F. Yu and Y.S. Wu, ``Non-linear $\hat{W_\infty}$
Current Algebra in the
$SL(2,R)/U(1)$ Coset Model,'' preprint UUHEP-91-19,
``Nonlinearly Deformed $W_\infty$
Algebra and Second Hamiltonian Structure of KP Hierarchy,
preprint UUHEP-91-09;
\item{} I. Bakas and E.B. Kiritsis, ``Beyond the Large $N$ Limit:
Nonlinear
$W_\infty$ as Symmetry of the $SL(2,R)/U(1)$ Coset Model,''
preprint UCB-PTH-91/44,
LBL-31213, UMD-PP-92-37.

\item{[12]} C.N. Pope, L.J. Romans and X. Shen, Phys. Lett. {\bf B236}
(1990) 173.

\item{[13]} C.N. Pope, L.J. Romans and X. Shen, Nucl. Phys. {\bf B339}
(1990) 191.

\item{[14]} C.N. Pope, L.J. Romans and X. Shen, Phys. Lett. {\bf B242}
(1990) 401.

\item{[15]} T. Inami, Y. Matsuo and I. Yamanaka, Phys. Lett. {\bf B215}
(1988) 701;
\item{} K. Honefeck and E. Ragoucy, Nucl. Phys. {\bf B340} (1990) 225;
\item{} K. Honefeck, Phys. Lett. {\bf B252} (1990) 357;
\item{} H. Lu, C.N. Pope, L.J. Romans, X.Shen and X.J. Wang, Phys. Lett.
{\bf B264} (1991) 91
; \item{} J. Evans and T. Hollowood, Nucl. Phys. {\bf B352} (1991) 723;
\item{} K. Ito, Phys. Lett. {\bf B259} (1991) 73;
\item{} H. Nohara and K. Mohri, Nucl. Phys. {\bf B349} (1991) 253;
\item{} D. Nemeschansky and S. Yankielowicz, ``$N=2$ $W$-algebras,
Kazama-Suzuki
Models and Drinfel'd-Sokolov reduction,'' USC-91/005 (1991);
\item{} L.J. Romans, ``The $N=2$ Super-$W_3$ Algebra,'' preprint USC-91/HEP06.

\item{[16]} L. Slater, ``Generalized Hypergeometric Functions,'' Cambridge
University Press (1966).

\item{[17]} C.N. Pope, L.J. Romans and X. Shen, Phys. Lett. {\bf B245}
(1990) 72.

\item{[18]} J. Hoppe, MIT Ph.D. Thesis (1982).

\item{[19]} E. Bergshoeff, M.P. Blencowe and K.S. Stelle, Comm. Math. Phys.
{\bf 128}
(1990) 213.

\item{[20]} I. Bakas, Phys. Lett. {\bf 228B} (1989) 57.

\item{[21]} I. Bakas, in {\sl Supermembranes and Physics in $2+1$
Dimensions}, Ed. M. Duff {\sl et al} (World Scientific, 1990).

\item{[22]} C.N. Pope and X. Shen, Phys. Lett. {\bf 236B} (1990) 21.

\item{[23]} A.O. Radul and I.D. Vaysurd, ``Differential Operators and
$W$ Algebras,''
preprint RI-09-91.

\item{[24]} E. Witten, Comm. Math. Phys. {\bf 113} (1988) 529.

\item{[25]} I. Bakas and E. Kiritsis, Nucl. Phys. {\bf B343} (1990) 185.

\item{[26]} E. Bergshoeff, C.N. Pope, L.J. Romans, E. Sezgin and X. Shen,
Phys. Lett. {\bf B245} (1990) 447.

\item{[27]} D.A. Deprieux, Phys. Lett. {\bf B252} (1990) 586.

\item{[28]} M. Fukuma, H. Kawai and R. Nakayama, ``Infinite-dimensional
Grassmannian
Structure of Two-dimensional Quantum Gravity,'' UT-572-TOKYO, Nov. 1990.

\item{[29]} X. Shen and X.J. Wang, ``Bosonization of the Complex Boson
Realizatio of $W_\infty$,'' CTP-TAMU-59-91, to appear in Phys. Lett. {\bf B}.

\item{[30]} C.M. Hull, Phys. Lett. {\bf B240} (1989) 110.

\item{[31]} K. Schoutens, A. Sevrin and P. van Nieuwenhuizen, Phys. Lett.
{\bf B243} (1990) 245.

\item{[32]} E. Bergshoeff, C.N. Pope, L.J. Romans, E. Sezgin, X. Shen and
K.S. Stelle, Phys. Lett. {\bf B243} (1990) 350.

\item{[33]} E. Bergshoeff, C.N. Pope, L.J. Romans, E. Sezgin and X. Shen,
Mod. Phys. Lett. {\bf A5} (1990) 1957.

\item{[34]} E. Bergshoeff, P. Howe, C.N. Pope, L.J. Romans, E. Sezgin,
X. Shen and K.S.
Stelle, ``Quantization Deforms $w_\infty$ to $W_\infty$ Gravity,''
CTP-TAMU-25/91.

\item{[35]} C.N. Pope, X.Shen, K.W. Xu and K.J. Yuan,
``$SL(\infty,R)$ Symmetry of
Quantum $W_\infty$ Gravity,'' CTP-TAMU-37-91, to appear in Nucl. Phys.
{\bf B}.

\item{[36]} F. David, Mod. Phys. Lett. {\bf A3} (1988) 1651;
\item{} J. Distler and H. Kawai, Nucl. Phys. {\bf B321} (1989) 509.

\item{[37]} K. Yamagishi, ``${\widehat W}_\infty$ Algebra Is  Anomaly
Free at $c=-2$,'' preprint, LLNL/HEP05 (1990).

\item{[38]} C.N. Pope, L.J. Romans and X. Shen, Phys. Lett. {\bf B254}
(1991) 401.

\item{[39]} D. Friedan, E. Martinec and S. Shenker, Nucl. Phys. {\bf B271}
(1986) 93.

\item{[40]} J. Distler, Nucl. Phys. {\bf B342} (1990) 523.

\item{[41]} A.M. Polyakov, Mod. Phys. Lett. {\bf A2} (1987) 893.

\item{[42]} V. Knizhnik, A.M. Polyakov and A.B. Zamolodchikov,
Mod. Phys. Lett. {\bf
A3} (1988) 819.

\item{[43]} V. Knizhnik and A.B. Zamolodchikov, Nucl. Phys. {\bf B247}
(1984) 83.

\item{[44]} E. Brezin and V. Kazakov, Phys. Lett. {\bf B236} (1990) 144;
\item{} D.J. Gross and A.A. Migdal, Phys. Rev. Lett. {\bf 64} (1990) 717;
\item{} M. Douglas and S. Shenkar, Nucl. Phys. {\bf B335} (1990) 635.

\item{[45]}  E. Witten, Nucl. Phys. {\bf B340} (1990) 281;
\item{} R. Dijkgraaf and E. Witten, Nucl. Phys. {\bf B342} (1990) 486;
\item{} K. Li, Nucl. Phys. {\bf B354} (1991) 725.

\item{[46]} H. Verlinde and E. Verlinde, Nucl. Phys. {\bf B348} (1991) 457.

\item{[47]} M. Goulian and M. Li, Phys. Rev. Lett. {\bf 66} (1991) 2051;
\item{} Vl.S. Dotsenko, Mod. Phys. Lett. {\bf A6} (1991) 3601;
\item{} Y. Kitazawa, Phys. Lett. {\bf 265B} (1991) 262;
\item{} N. Sakai and Y. Tanii, Prog. Theor. Phys. {\bf 86} (1991) 547.

\item{[48]} S.R. Das and A. Jevicki, Mod. Phys. Lett. A5 (1990) 1639.

\item{[49]} M. Douglas, Phys. Lett. {\bf B238} (1990) 176;
\item{} D.Gross and A. Migdal, Nucl. Phys. {\bf B340} (1990) 333.

\item{[50]} A.M. Polyakov, Mod. Phys. Lett. {\bf 6A} (1991) 635.

\item{[51]} I. Klebanov, D. Gross and M. Newman, Nucl. Phys. {\bf B350}
(1991) 621.
\item{} U.H. Danielsson and D. Gross, Nucl. Phys. {\bf B366} (1991) 3.

\item{[52]} E. Witten and B. Zwiebach, ``Algebraic Structure and Differential
Geometry in 2D String Theory,'' preprint IASSNS-HEP-92/4, MIT-CTP-2057.

\item{[53]} S. Odake and T. Sano, Phys. Lett. {\bf B258} (1991) 369.

\item{[54]} E. Witten, Comm. Math. Phys. {\bf 117} (1988) 353.

\item{[55]} E. Witten, ``Surprises with Topological Theories,''
Contribution to the
String '90 at Texas A\&M, 1990;
\item{} K. Li, Nucl. Phys. {\bf B346} (19900 329, Phys. Lett. {\bf B251}
(1990) 54;
\item{} H. Lu, C.N. Pope and X. Shen, Nucl. Phys. {\bf B366} (1991) 95.

\item{[56]} C.M. Hull, ``Classical and Quantum $W$ Gravity,''
preprint QMW/PH/92/1.

\item{[57]} H. Lu, C.N. Pope, L.J. Romans, X. Shen and X.J. Wang,
Phys. Lett. {\bf
B267} (1991) 356.

\item{[58]} V.S. Dotsenko, ``Remarks on the Physical States and the
Chiral Algebra of 2D
Gravity Coupled to $c\le1$ Matter,'' preprint PAR-LPTHE 92-4.

\item{[59]} I. Bakas, B. Khesin and E. Kiritsis, preprint LBL-31303 (1991).

\end